\documentclass[twocolumn]{aastex631}

\usepackage{amsmath}
\usepackage{xcolor}

\usepackage{soul}

\begin{document}

\title{The coronal power spectrum from MHD mode conversion above sunspots}


\author[0000-0002-2235-3216]{Hemanthi Miriyala}
\affiliation{Department of Mathematics, Physics and Electrical Engineering, Northumbria University, Newcastle upon Tyne, NE1 8ST, UK}

\author[0000-0001-5678-9002]{Richard J. Morton}
\affiliation{Department of Mathematics, Physics and Electrical Engineering, Northumbria University, Newcastle upon Tyne, NE1 8ST, UK}

\author[0000-0003-3812-620X]{Elena Khomenko}
\affiliation{Instituto de Astrof\'{i}sica de Canarias, 38205 La Laguna, Tenerife, Spain}
 \affiliation{Departamento de Astrof\'{i}sica, Universidad de La Laguna, 38205, La Laguna, Tenerife, Spain}

\author[0000-0003-1529-4681]{Patrick Antolin}
\affiliation{Department of Mathematics, Physics and Electrical Engineering, Northumbria University, Newcastle upon Tyne, NE1 8ST, UK}

\author[0000-0002-5915-697X]{Gert J.J. Botha}
\affiliation{Department of Mathematics, Physics and Electrical Engineering, Northumbria University, Newcastle upon Tyne, NE1 8ST, UK}

\begin{abstract}
Sunspots are intense regions of magnetic flux that are rooted deep below the photosphere. It is well established that sunspots host magnetohydrodynamic waves, with numerous observations showing a connection to the internal acoustic (or \textit{p-})modes of the Sun. The \textit{p-}modes are fast waves below the equipartition layer and are thought to undergo a double mode conversion as they propagate upwards into the atmosphere of sunspots, which can generate Alfv\'{e}nic modes in the upper atmosphere. We employ 2.5D magnetohydrodynamics (MHD) numerical simulations to investigate the adiabatic wave propagation and examine the resulting power spectra of coronal Alfv\'{e}nic waves. A broadband wave source is used that has a 1D power spectrum which mimics aspects of the observed \textit{p-}mode power spectrum. We examine magnetoacoustic wave propagation and mode conversion from the photosphere to the corona. Frequency filtering of the upwardly propagating acoustic waves is a natural consequence of a gravitationally stratified atmosphere, and plays a key role in shaping the power spectra of mode converted waves. We demonstrate that the slow, fast magnetoacoustic waves and Alfv\'{e}n waves above the equipartition layer have similarly shaped power spectra, which are modified versions of the driver spectrum. Notably, the results reveal that the coronal wave power spectra have a peak at a higher frequency than that of the underlying \textit{p}-mode driver. This matches observations of coronal Alfv\'enic waves and further supports the role of mode conversion process as a mechanism for Alfv\'enic wave generation in the Sun's atmosphere.

\end{abstract}


\section{Introduction} \label{sec:intro}
Magnetohydrodynamic (MHD) waves are considered to play a key role in the transfer of energy through the Sun's atmosphere \citep{Osterbrock1961,Jess2015,Nakariakov_2020,VanD_2020}, carrying energy from the convective motions in the photosphere out into the corona and beyond.
In an inhomogeneous plasma, such as the Sun's atmosphere, a variety of magnetohydrodynamic (MHD) wave modes can exist beyond the traditional slow, fast and Alfv\'en modes \citep{Spruit1982,EdwinRoberts1983}. Inhomogeneity perpendicular to the magnetic field leads to MHD waves having mixed properties \citep{Goossens_2019}. As such, a variety of modes can be considered Alfv\'enic in nature \citep{goossens2009nature}. Their characteristic incompressibility indicate that the Alfv\'{e}nic waves play a crucial role in transporting energy through the solar atmosphere \citep[e.g.,][]{Morton2022}.

Alfv\'{e}nic modes are generally considered to be driven by the buffeting of magnetic fields in the photosphere \citep[e.g.,][]{Cranmer2005}. However, global observations of coronal Doppler velocities suggest Alfv\'{e}nic waves have an enhanced power around 4~mHz \citep{morton2019}, which is not expected from convective driving. It has been suggested that the enhanced power can be linked with the peak of the \textit{p-}mode power spectrum in the photosphere, which lies at $\sim3$~mHz \citep[e.g.,][]{Hansen_2012}. The close proximity of the peaks in frequency space has been taken as evidence that the coronal Alfv\'{e}nic waves are somehow influenced by the \textit{p}-modes.

Previous studies have demonstrated the possibility of converting acoustic modes to Alfv\'en waves \citep[via conversion to and from the fast magnetoacoustic mode, e.g.,][]{Cally2008, Khomenko2012, felipe2012three}. A number of other studies have also implemented a broadband \textit{p}-mode driver in order to excite coronal Alfv\'{e}nic waves. For instance, \citet{kuniyoshi_2024} used 2D simulations to demonstrate high-frequency transverse spicule oscillations driven by \textit{p}-modes, although the underlying mechanism exciting these oscillations remains unclear. \citet{gao2023modeling} utilize a 3D model of a closed magnetic loop and excite waves with a mono-periodic \textit{p}-mode driver. There is certainly the possibility of linear mode conversion in such simulations, but their focus is on the generation of standing waves. 
\citet{skirvin_2023} investigated the mechanism for exciting transverse Alfv\'enic waves using an inclined \textit{p}-mode wave driver, which breaks symmetry and utilises pressure to displace the magnetic field. However, this study did not explicitly address mode conversion. Related work by \citet{skirvin} explored the role of transverse structuring in mode conversion within the lower solar atmosphere. Despite such studies, it has not yet been demonstrated that enhancements in coronal power spectra can arise from  \textit{p}-mode excitation of Alfv\'enic waves.
\\

The \textit{p-}modes are the pressure perturbations trapped below the photosphere \citep{Takashi} and are absorbed by regions of high magnetic fields, such as sunspots or magnetic bright points associated with the network regions in the quiet Sun. Due to the abundant magnetic field, the \textit{p-}modes are funnelled as magnetoacoustic waves into the solar atmosphere \citep{Spruit1992, Cally1997}. The \textit{p-}modes are predominantly acoustic in nature and are subject to the acoustic cut-off frequency. The frequency of the cutoff arising in a gravitationally stratified plasma is (from a WKB approximation) given by 
	\begin{equation}\label{eq:cutoff}
	    \nu_{\rm ac} =\frac{\gamma\textit{g}}{4\pi c_{s}},
    \end{equation}
where $c_{s}$ is the sound speed, $\gamma$ is the ratio of specific heats, and g is gravity \citep{Landgraf1997,2011, Khomenko2011}. Acoustic (fast) modes in a high-beta plasma (i.e., in the low photosphere and solar interior) propagate isotropically, hence are little influenced by the magnetic field. Although when the Alfv\'en speed, $v_A$, and sound speed are comparable ($c_s\approx v_A$), then the magnetic field can influence wave propagation \citep{Cally_2006}. 

In a low-beta plasma ($c_s<v_A$) the acoustic waves are the slow modes and are field-guided. Hence the cut-off frequency is modified by effective gravity along the inclined flux tubes as the slow magnetoacoustic waves have a preferred path of propagation dictated by the inclined magnetic field  \citep[e.g.,][]{Schunker2006}. The effective cut-off is
$$
\nu_{\rm ac,eff} = \nu_{\rm ac}\cos\theta,
$$
here the cosine of the inclination angle, $\theta$, is defined with respect to the local vertical. The influence of the effective acoustic cut-off on the slow modes is thought to be the basis of well known phenomena associated with sunspot oscillations. One is that sunspot's umbrae show a power spectra dominated by oscillations with frequencies of $\sim3$~mHz in the photosphere but is dominated by $\sim5$~mHz oscillations in the chromosphere \citep[see, e.g., ][]{Bogden_2006,Centeno_2006, Felipe2010}. Moreover, the variation of peak oscillatory power with inclination has also been reported in the observations of sunspot's penumbral chromosphere. The frequency of slow magnetoacoustic waves with the largest power decreases  with distance from the spot centre \citep{jess2013influence, Jess2016, Morton_2021}. 

\medskip

\citet{Cally2008} first discussed mode conversion as a mechanism for producing Alfv\'{e}n waves from \textit{p-}modes. They demonstrated that Alfv\'{e}n waves can be generated by the mode conversion of fast magnetoacoustic waves when the magnetic field is inclined with respect to the plane of wave propagation. Motivated by these studies, \citet{Khomenko2011, Khomenko2012} employed 2.5D numerical simulations in sunspot-like regions to understand the efficiency of conversion from \textit{p-}modes to Alfv\'{e}n waves. The \textit{p-}modes are fast acoustic waves below the equipartition layer (the layer where $c_s=v_A$). The fast acoustic waves largely enter the low-beta atmosphere as fast magnetic waves, with mode conversion changing their character from acoustic to magnetic. However, around the equipartition layer, the fast acoustic waves can also be transmitted as slow magnetoacoustic modes for a narrow range of magnetic field inclinations \citep[e.g.,][]{Cally_2006, Schunker2006}. 

The fast magnetoacoustic waves then undergo significant reflection due to the rapidly increasing Alfv\'{e}n speed in the upper atmosphere at the locations where their horizontal phase speed matches the local Alfv\'{e}n speed i.e. $\omega/k_h = v_A$, where $\omega$ is the angular frequency and  $k_h$ is the horizontal wavenumber \citep{Cally2011}. The fast-to-Alfv\'{e}n conversion coefficient is then predominantly based on the horizontal wave number ($k_h$), magnetic field inclination ($\theta$) from the stratification direction, and the azimuthal angle ($\phi$) of the wave vector with respect to the plane containing the stratification and magnetic field directions \citep{Cally2011}. Hence there is effectively a double mode conversion in getting from acoustic to Alfv\'{e}n waves. The extent of acoustic to Alfv\'{e}n conversion is largely influenced by the magnetic field inclination and azimuth angles around the equipartition layer \citep{Cally2008, Khomenko2012}. \citet{Cally2008} reported that magnetic field inclinations between $30^\circ - 40^\circ$ and azimuth angles between $60^\circ - 80^\circ$ at the equipartition layer favour the double mode conversion, and the resultant Alfv\'{e}n fluxes are significantly higher than the acoustic fluxes.

\medskip

Given the nature of MHD wave propagation in the lower solar atmosphere, how then might the observed enhancement of coronal Alfv\'enic wave power occur? And why is the peak at a frequency of 4~mHz while the \textit{p}-modes peak around 3.3~mHz? We suggest the shape of the coronal Alfv\'enic power spectrum could be defined by the atmospheric filtering of the \textit{p}-mode power spectrum. The acoustic cut-off frequency is able to modify the acoustic power spectrum through a frequency filtering of the upwardly propagating acoustic modes.

If we consider the acoustic spectrum at the equipartition layer, the effect of the cut-off will be to skew the peak of the power to higher frequencies than that of the \textit{p}-mode spectrum, as the lower frequencies are truncated. The mode conversions from fast (acoustic) to fast (magnetic) and fast (magnetic) to Alfv\'en are linear; hence there is no change in wave frequency. As such, one should also expect the power spectra of fast and Alfv\'en modes, generated by mode conversion from upwardly propagating acoustic modes, to have a peak frequency higher that of the \textit{p}-modes. The frequency filtering will depend on the height of the equipartition layer. The cut-off frequency varies as a function of height in the solar atmosphere, having its largest value at the temperature minimum. If the equipartition layer occurs below the temperature minimum along a nearly vertical field line as in a sunspot umbra, then the peak frequency of the fast and Alfv\'en mode spectra will likely be lower than that of the slow modes. This is because the slow mode is continuosly influenced by the effect of the cut-off after propagating past the equipartition layer and there is further reflection due to the transition region. 

\medskip

In the following, we examine the role the acoustic cut-off on MHD wave propagation with numerical simulations. Previous work on the fast-to-Alfv\'en mode conversion focused efforts on understanding the fundamentals of the process, generally opting to use monochromatic wave drivers for clarity \citep{Khomenko2011, Khomenko2012}. A non-monochromatic driver was used by \cite{Felipe2010}, but their simulations did not reach the corona and the power spectra of the Alfv\'enic waves did not appear to be of interest. As such, there has not been an investigation into what aspects of the \textit{p}-mode spectrum are imparted upon the coronal Alfv\'enic waves. Hence, the main objective of this work is to examine the nature of the coronal Alfv\'{e}nic wave power spectrum when the system is driven by a broadband driver that resembles the \textit{p-}modes. We employ a modified version of the sunspot model used in the previous studies of \cite{Khomenko2011, Khomenko2012} and extend the atmosphere into the transition region and corona. Acoustic modes are driven with a broadband driver and we investigate the adiabatic wave propagation, examining the resulting power spectra of coronal Alfv\'{e}n waves. 

\section{Numerical Setup}

\begin{figure*}[t]
    \centering
        \centering        \includegraphics[width=0.9\linewidth]{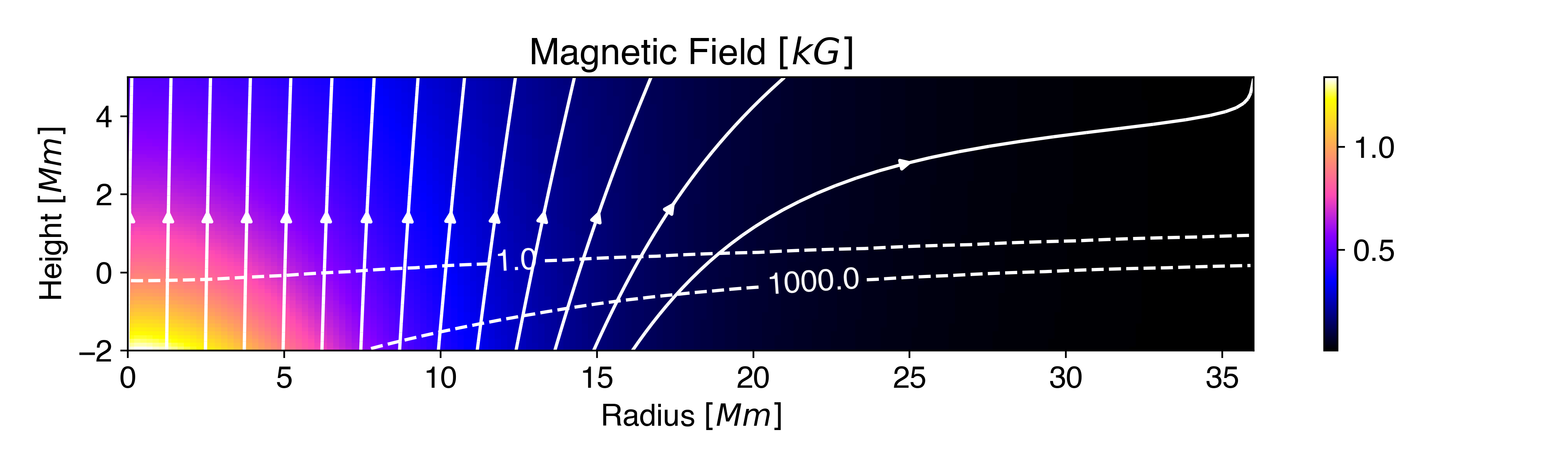}
        \centering       \includegraphics[width=0.9\linewidth]{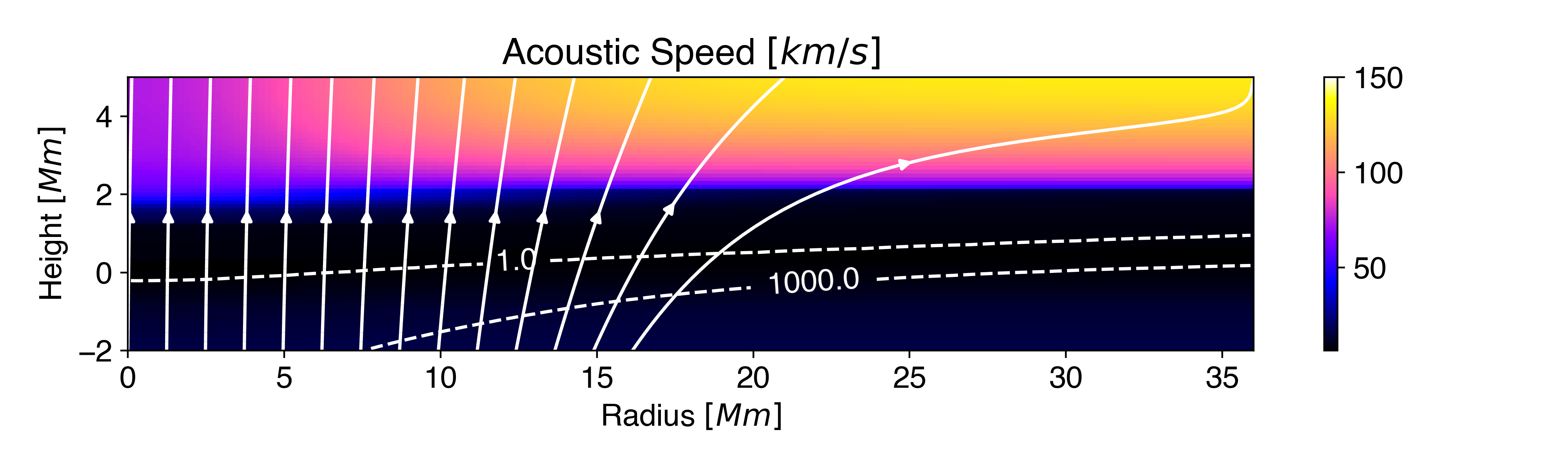}
        \centering        \includegraphics[width=0.9\linewidth]{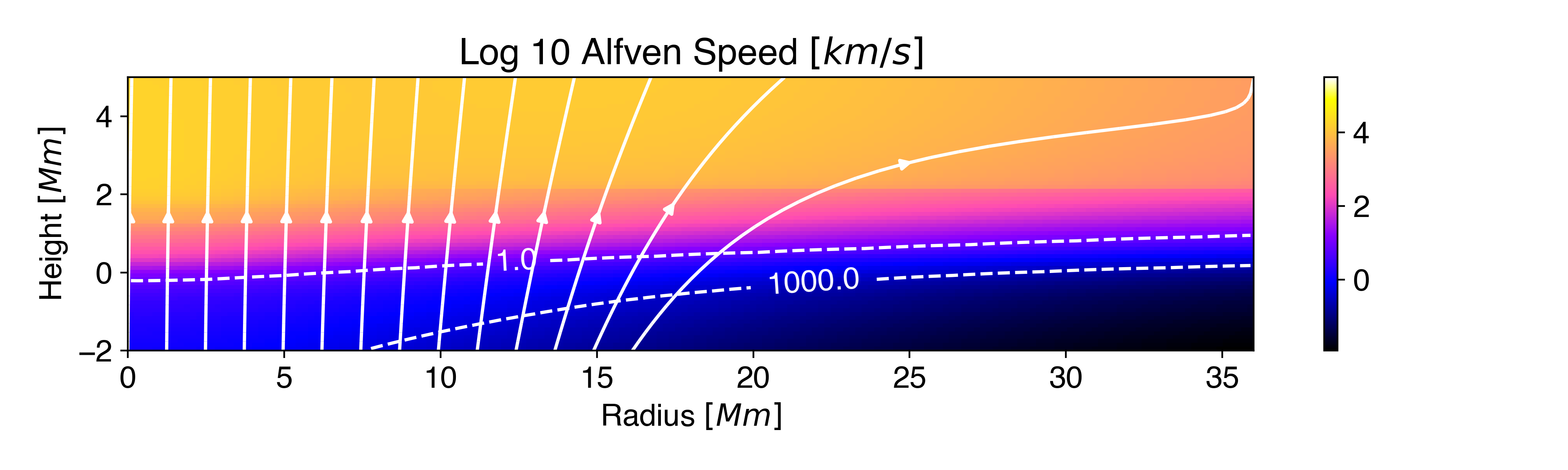}
    
    \caption{Topology of the expanded sunspot solution that includes transition region and corona. Following \citet{Khomenko2008}, their equations 6 and 7, the model is constructed using the following parameters:  $a = 2h$, h = 3~Mm, $B_0 = 20000$ G, $z_0 = 1$ Mm, and $\eta = 3.5$. White lines are magnetic field lines. Dashed lines with labels are the contours of the ratio of the sound speed and the Alfv\'{e}n speed squared, $(c_s^2/v_A^2)$.}
    \label{fig:2Dcylinder}
\end{figure*}

\begin{figure*}[t]
    \centering
    \resizebox{\hsize}{!}{
    \includegraphics[scale=1.5]{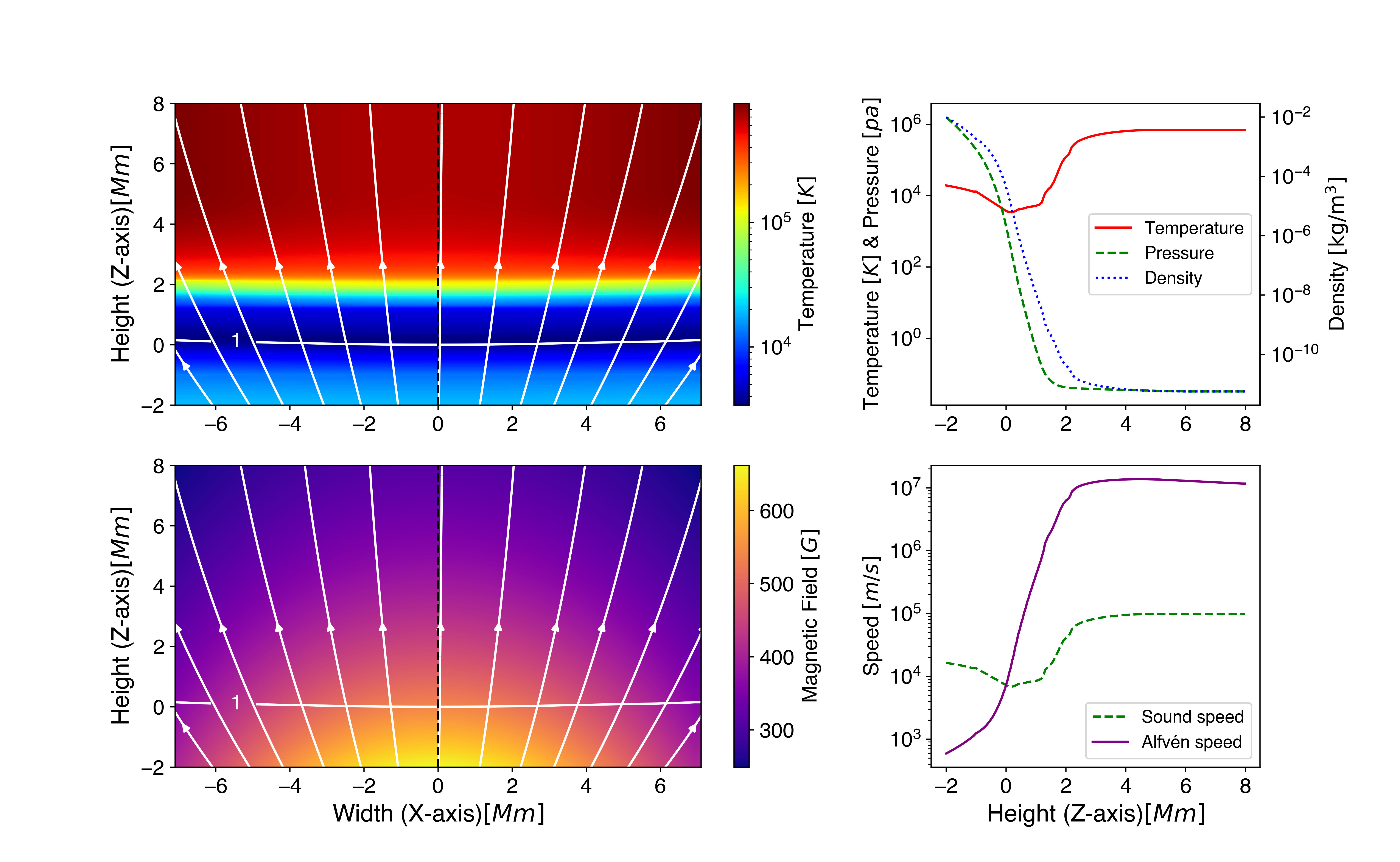}
    }
    \large
    \caption{Background atmosphere taken at Y=7~Mm (from the Y-origin at the spot centre). The left panels in the top and bottom are the temperature and magnetic field variations of the background atmospheric conditions, respectively. The white contours with arrows are the in-plane projections of the background magnetic field lines and the near-horizontal curve across the domain is the equipartition layer where $c_s/v_A=1$. The top right panel shows the plasma properties as a function of height at the centre of the domain (corresponding to the vertical dashed line in the left panels). Similarly, the bottom right panel is the variation of sound and Alfv\'{e}n speeds as a function of height at the centre of the domain.}
    \label{fig:background_parameters}
\end{figure*}

\subsection{Numerical Scheme} \label{subsec:numericalscheme}
Following \citet{Khomenko2011, Khomenko2012}, we use the MANCHA code to solve the non-linear equations for perturbations where the equilibrium state is removed from the equations \citep[see][for more details on MANCHA]{Khomenko2006,Khomenko2008, Felipe2010, Modestov2024}. The system of MHD equations to be solved are written in conservative form, namely,
\\

\begin{equation}
    \frac{\partial \rho}{\partial t} + \mathbf{\nabla} \cdot (\rho \mathbf{v}) = 0,
        \label{eq:mass_conv}
\end{equation}

\begin{equation}
    \frac{\partial(\rho\mathbf{v})}{\partial t} + \mathbf{\nabla} \cdot \left[\rho\mathbf{v}\mathbf{v} + \left(p + \frac{\mathbf{B^2}}{2\mu_0}
    \right)\mathbf{I} - \frac{\mathbf{BB}}{\mu_0}\right] = \rho \mathbf{g},
\end{equation}
\begin{equation}
\end{equation}
\begin{equation}
    \frac{\partial \mathbf{B}}{\partial t} = \mathbf{\nabla} \mathbf{\times} (\mathbf{v} \times \mathbf{B})
\end{equation}
where $\mathbf{I}$ is the Identity tensor and $E$ represents the total energy and is expressed as
\begin{equation}
    E = \frac{1}{2} \rho v^2 + \frac{p}{\gamma - 1} + \frac{\mathbf{B}^2}{2\mu_0}.
    \label{eq:E}
\end{equation}
 Here, $\rho$ is the density, \textbf{v} is the velocity, p is the gas pressure, \textbf{B} is the magnetic field, $\mu_0$ the magnetic permeability, \textbf{g} is the gravitational acceleration. We also employ an ideal equation of state for which $\gamma=5/3$. The MHD equations are solved with spatial and temporal discretisations that are centered, fourth-order accurate, explicit finite differences scheme \citep{vogler} and fourth order Runge-Kutta scheme respectively \citep{Khomenko2006, Modestov2024}.
Following \citet{Khomenko2012}, we use a 2.5D approximation to solve the equations, which
allows for vectors in three spatial directions, but the derivatives are taken only in two directions (one vertical and one horizontal). Hence the perturbations only propagate in the $XZ$ plane.

\subsection{Magneto-static Sunspot Model} \label{subsec:sunspotmodel}
To generate the background atmosphere upon which the wave propagation occurs, we choose to sample a 2D slice from a 3D atmosphere. For this purpose, we employ a sunspot model that closely resembles the one discussed in \citet{Khomenko2008}. The sunspot domain is a thick flux tube which is azimuthally symmetric with no twist. It is a current distributed model which has the radial variations of field strength and gas pressure continuous across the spot, and it is constructed by merging a self-similar solution by \citet{low1980exact} in the deep layers with the model of \citet{Pizzo1986} in the atmospheric layers. There is no sharp transition between the umbra and penumbra or between the penumbra and the field-free photosphere. The magnetic field inclination of the field lines changes gradually from the sunspot axis outward (see Figure~\ref{fig:2Dcylinder}). At the spot centre, $(X,Y)=(0,0)$~Mm, the magnetic field is 2200~G below the photosphere and gradually decreases with height. Readers are referred to \cite{Khomenko2008} for details on the construction of the sunspot model.

\medskip
There are a number of modifications between our model and that of \cite{Khomenko2008}. One adaption is that we elect to use the FAL-C model \citep{Fontenla1993} as our quiet Sun boundary (between 0.7~Mm and 2.2~Mm in height), which describes the upper photosphere and solar chromosphere \citep[FAL-C is more consistent with the observed hydrogen and helium spectra than VAL-C;][]{Fontenla1993}.

In order to include a corona in the simulation domain, we choose to extend the atmosphere until 8~Mm above the photosphere. The upper 2~Mm of the domain are reserved for accommodating the boundary conditions at the top of the simulation domain. To extend the atmosphere of the quiet Sun, we first interpolate the temperature profile using a polynomial function that begins at 2.2~Mm and attains a constant temperature by 5.5~Mm. We extend the atmosphere of the spot centre likewise. Next, we calculate pressure and density assuming hydrostatic equilibrium as described in \citet{santamaria2015magnetohydrodynamic}. First, the pressure scale height is calculated as,
    \begin{equation}\label{eq:scale_height}
        H_p = \frac{R_{\rm gas}T}{g\mu_{\rm var}},
    \end{equation}
    which is used in the solution for the following hydrostatic equilibrium for pressure
    \begin{equation}\label{eq:pressure}
        \frac{dp(z)}{dz} + \frac{p}{H_p} = 0.
    \end{equation}
Finally, we recover the density distribution from equations (\ref{eq:scale_height}) and (\ref{eq:pressure}) using
    \begin{equation}
        \rho = \frac{p}{gH_p}.
    \end{equation}

\noindent Here, $p$ is the pressure, $\rho$ is the density, $R_{\rm gas}$ is the gas constant, and $T$ is the temperature, $z$ is the height. We expect a varying degree of ionisation of plasma with height. The mean atomic weight ($\mu_{\rm var}$) is approximately 0.5 in a single fluid hydrogen-only plasma. Hence, the value of $\mu_{\rm var}$ is smoothly decreased until it reaches 0.5 in the corona for both spot centre and quiet sun. We achieve this extrapolation past 2.2~Mm using an exponential function:
   \begin{equation}
       \mu_{\rm var} = 0.5 + e^{\kappa(z_i - z)}
   \end{equation}
 where, $\kappa$ is a scaling factor or sometimes referred to as steepness parameter, $z_i$ is the initial height at which we begin the extrapolation. Hence, we extrapolate 1D atmospheric profiles past 2.2~Mm respectively, for both quiet sun and spot centre. Once these models are established, 
smooth transition between them for the gas pressure and
scale height distributions is achieved and the force balance equation along
the magnetic field lines is iterated until
a convergence criterion is reached \citep{Pizzo1986, Khomenko2008}.

Figure \ref{fig:2Dcylinder} displays the atmosphere at the centre of the sunspot ($Y=0$) cropped from -2~Mm to 5~Mm in Z-direction and from 0~Mm to 38~Mm in the radial direction for visualization purposes. Figure \ref{fig:2Dcylinder} clearly shows the variations across the domain in both vertical and radial directions. The umbral and penumbral region in the model can be differentiated based on the inclinations of magnetic field lines.

\medskip

For our 2.5D simulation, we use a vertical slice ($XZ$ plane) located at a distance of 7~Mm away from the centre of the sunspot in the Y-direction. The domain is also restricted to 14.2~Mm in $X$ (and is 10~Mm in $Z$ direction). The magnetic structure of magnetostatic solution is shown in the bottom left panel of Figure~\ref{fig:background_parameters}, which is a plot of magnetic field strength with in-plane projections of the background magnetic field lines for the sunspot on an extended vertical scale of 10~Mm (i.e. [-2,8]~Mm). The temperature of the background atmosphere is shown in the top left panel of Figure~\ref{fig:background_parameters}. An example of the 1D plasma profiles from the model can be observed in the right panels of Figure~\ref{fig:background_parameters},  located at $(X,Y)=(0,7)$~Mm  (location indicated by the the vertical dashed line in the left panels of the Figure~\ref{fig:background_parameters}). 
The spatial resolution across the domain is uniform, and is set to 50~km in the horizontal $X$-direction and 20~km in the vertical $Z$-direction.

\subsection{Diffusion Profile and Boundary Conditions} \label{subsec:boundary}

For the model boundary conditions in the horizontal direction, we follow \cite{MacBride2022}. We use periodic boundary conditions on either side on our simulation domain by reflecting the model horizontally and then shifting the model by half the original width such that the original domain remains in the center of the $X$-axis.  The numerical domain is large enough such that we do not see wave entering back from the outer edges.
Despite having our driver exciting perturbations with small amplitudes, they undergo appreciable amplification with height due to stratification. In order to reduce reflections from the top boundary layer, we introduce a layer of diffusion on the top boundary above 4~Mm until 8~Mm. The diffusion profile is constructed using a sigmoid function given by,
\begin{equation}
    D =  1/(1 + {e^{\kappa(z-z_{c})}}).
\end{equation}

\begin{figure}[t]
    \centering
    {
    \includegraphics[scale=0.55, trim=0 0 0 0, clip]{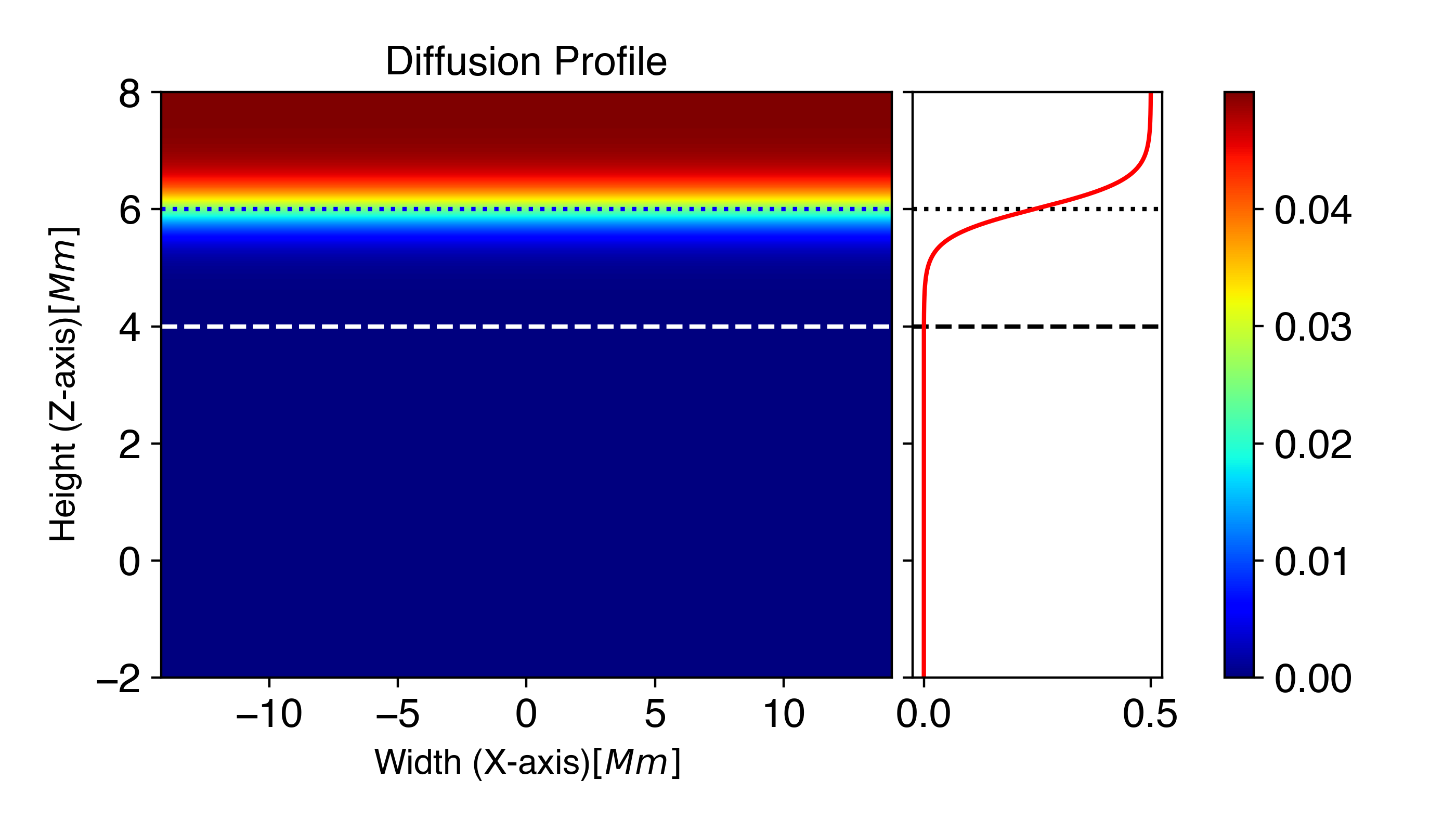}
    }      
    \caption{2D and 1D plots of the diffusion profile. The dotted lines indicate the height at which the sigmoid is centred ($z_{c}$). The dashed lines indicate the height we consider to compute the Alfv\'{e}n power spectrum.}
    \label{diff}
\end{figure}

\noindent Here, $\kappa$ is a scaling factor, $z$ is the height, $z_{c}$ is the height at which the sigmoid is centered. From Figure~\ref{diff} it can be seen that the sigmoid starts after 4~Mm and is centred at around 6~Mm. We use the 2D diffusion profile discussed above as a mask which multiplies time-constant part of the diffusion coefficient (proportional to the sum of the flow speeds and the grid spacing in each Cartesian direction). The final diffusion coefficient, different for each equation and direction, is formed by the sum of the time-constant part, hyperdiffusion and shock diffusion contributions, computed as explained in section 3.4 in \citet{Modestov2024}. The diffusion coefficients then enter into the computations of the viscosity tensor, Ohmic diffusion, their corresponding counterparts in the energy equation, as well an an artificial term in the continuity equation that does not have a physical counter-part. For more details, the reader is referred to \citet{Modestov2024}.

Additionally, PML with a sponge layer (SPML), is applied to the upper part of the model from 6~Mm until 8~Mm (100 grid points) as part of the boundary conditions \citep{Modestov2024}. PML has proved to be an excellent wave absorber and has been employed in many previous works \citep{Khomenko2008, Felipe2010, Khomenko2012, MacBride2022}. As the Alfv\'{e}n and sound speeds increase drastically with height in the corona, waves with large amplitudes develop in our simulation.

\subsection{Broadband Driver} \label{subsec:driver}
\begin{figure}[t]
\centering
    \resizebox{\hsize}{!}{
    \includegraphics[scale=0.08, trim=0 0 0 20, clip]{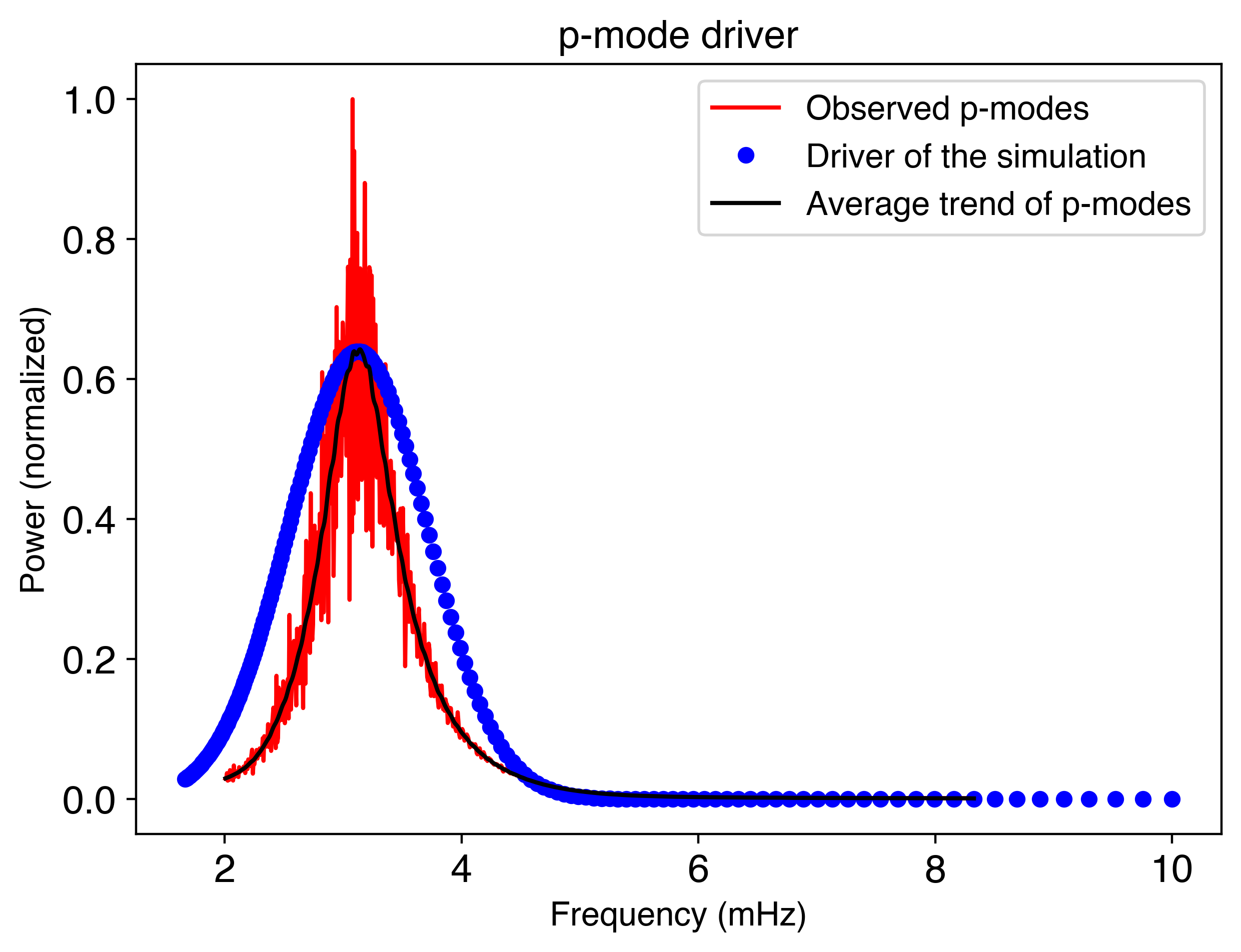}
    }
    \large
    \caption{The frequency dependence of the wave driver. A comparison of the broadband driver (blue dots) used in the simulation with the observed $p$-mode power spectrum (red solid) from SOHO MDI \citep{Rhodes1997}. The black line is the average trend fitted to the observed \textit{p}-mode power spectrum.}
    \label{driver}
\end{figure}

We employ a broadband driver designed to mimic aspects of the observed $p$-mode spectrum (which is shown in Figure~\ref{driver}). The feature of interest is the distinct peak at 3 mHz. The $p$-modes with frequencies much lower than this value are always evanescent in the atmosphere, so likely do not contribute to the flux of coronal Alfv\'en waves. Further, given that the mode conversion process is linear, we expect a one-to-one mapping between the $p$-mode frequencies and those of the coronal Alfv\'enic waves. Hence, we also do not attempt to simulate the high-frequency regime here. To describe our driver, we use a Gaussian function centered at $P_{c} = 320$~s ($\nu_c\approx3$~mHz to model the velocity amplitude in frequency space, i.e.,
\begin{equation}
V_n = v_0\exp\left(-\frac{1}{2}\left(\frac{\nu_n-\nu_c}{\sigma_\nu} \right)^2\right).
\label{eq:v_amp_driver}
\end{equation}
Here, $v_0 = 2 \times 10^{-4}$ m/s and $\sigma_{\nu}$ is the standard deviation of the Gaussian. The shape of the driver power spectrum is shown as the blue curve in Figure~\ref{driver}. For the driver, we consider 200 sinusoidal perturbations, with periods ($P_n=1/f_n$) uniformly spaced between 100 - 600 seconds, with the amplitude for each sinusoid given by Eq.~\ref{eq:v_amp_driver}.
           
The driver is confined vertically to a few grid points close to the domain's bottom boundary (Z= -2~Mm -1.25~Mm and at X=0~Mm). The form of the perturbations is determined analytically as an acoustic-gravity wave \citep[see][]{Mihalas1986,Khomenko2012}, ignoring the magnetic field and temperature gradient. The ratio of sound-to-Alfv\'en speeds squared in the driving region is $c_s^2/v_A^2\approx 250$. Hence, we can expect the magnetic field to be dynamically unimportant and mainly acoustic modes to be excited by the driver. In accordance with \citet{Mihalas1986}, self-consistent perturbations of the velocity vector, pressure, and density are given by:
    \begin{eqnarray}
		\delta V_{z} = \sum_{n=1}^{200}V_n&&g(x)\exp\left(\frac{z}{2H} + k_{zi}z\right) \nonumber \\
        && \times \sin(\omega_{n} t - k_{zr}z+\Phi_n) 
		\end{eqnarray}
		\begin{eqnarray}
		\frac{\delta p}{p_{0}} =\sum_{n=1}^{200}V_n|P_n|&&g(x)\exp\left(\frac{z}{2H} + k_{zi}z\right) \nonumber \\
        && \times\sin(\omega_{n} t - k_{zr}z + \phi_{P_n}+\Phi_n) 
		\end{eqnarray}
		\begin{eqnarray}
		\frac{\delta \rho}{\rho_{0}} =\sum_{n=1}^{200}V_n|R_n|&& g(x)  \exp\left(\frac{z}{2H} + k_{zi}z\right) \nonumber \\
        && \times\sin(\omega_{n} t - k_{zr}z + \phi_{R_n}+\Phi_n).
    \end{eqnarray}

\noindent Here, $H$ is the density scale height,  $k_{zr}$ and $k_{zi}$ are real and imaginary vertical wave numbers, and $\Phi_n$ is a random phase added to the wave at each of the 200 frequencies. The subscript 0 refers to quantities related to the unperturbed background atmosphere. The $X$-dependence of the pulse, denoted $g(x)$, is defined by:
\begin{equation}
    g(x) = \exp\left(-\frac{1}{2}\left(\frac{x-x_0}{\sigma_x} \right)^2\right),
\label{16}    
\end{equation}
where $\sigma_x$ defines the size of the pulse in $X$-direction, $x_0$ is the location where the Gaussian is centred and $x$ is the horizontal coordinate. We choose $\sigma_x=1.25$~Mm. Using a spatially localised pulse excites modes with different horizontal wavenumbers, with mode amplitude decreasing as the absolute value of the wave number increases \citep{Khomenko2006}. 

The amplitudes and the relative phase shifts between the perturbations are given by,
\begin{equation}
    |P_n| = \frac{\gamma}{\omega_{i_n}}\sqrt{k_{zr}^{2} + \left(k_{zi} + \frac{1}{2H}\frac{(\gamma - 2)}{\gamma}\right)^{2}},
\end{equation}
	
\begin{equation}
    |R_n| = \frac{1}{\omega_{i_n}}\sqrt{k_{zr}^{2} + \left(k_{zi} - \frac{1}{2H}\right)^{2}},
\end{equation}
	
\begin{equation}
    |\phi_{P_n}| = \arctan\left(\frac{k_{zi}}{k_{zr}} +\frac{1}{2Hk_{zr}} \frac{(\gamma - 2)}{\gamma}\right),
\end{equation}

\begin{equation}
    |\phi_{R_n}| = \arctan\left(\frac{k_{zi}}{k_{zr}} - \frac{1}{2Hk_{zr}} \right).
\end{equation}
	
\noindent Given the wave angular frequency, the vertical wavenumber is found from the dispersion relation for acoustic-gravity waves in an isothermal atmosphere as 
\begin{equation}
    k_{z} = k_{zr} + ik_{zi} = \sqrt{(\omega_{n}^{2}-\omega_{ac}^{2})/c_{s}^{2}}
\end{equation}
	
\noindent where $\omega_{ac}$ = $2\pi \nu_{ac}$ is the acoustic cutoff frequency. 

\begin{figure*}[t]
    \centering
    \resizebox{\hsize}{!}{
    \includegraphics[width=0.5\textwidth]{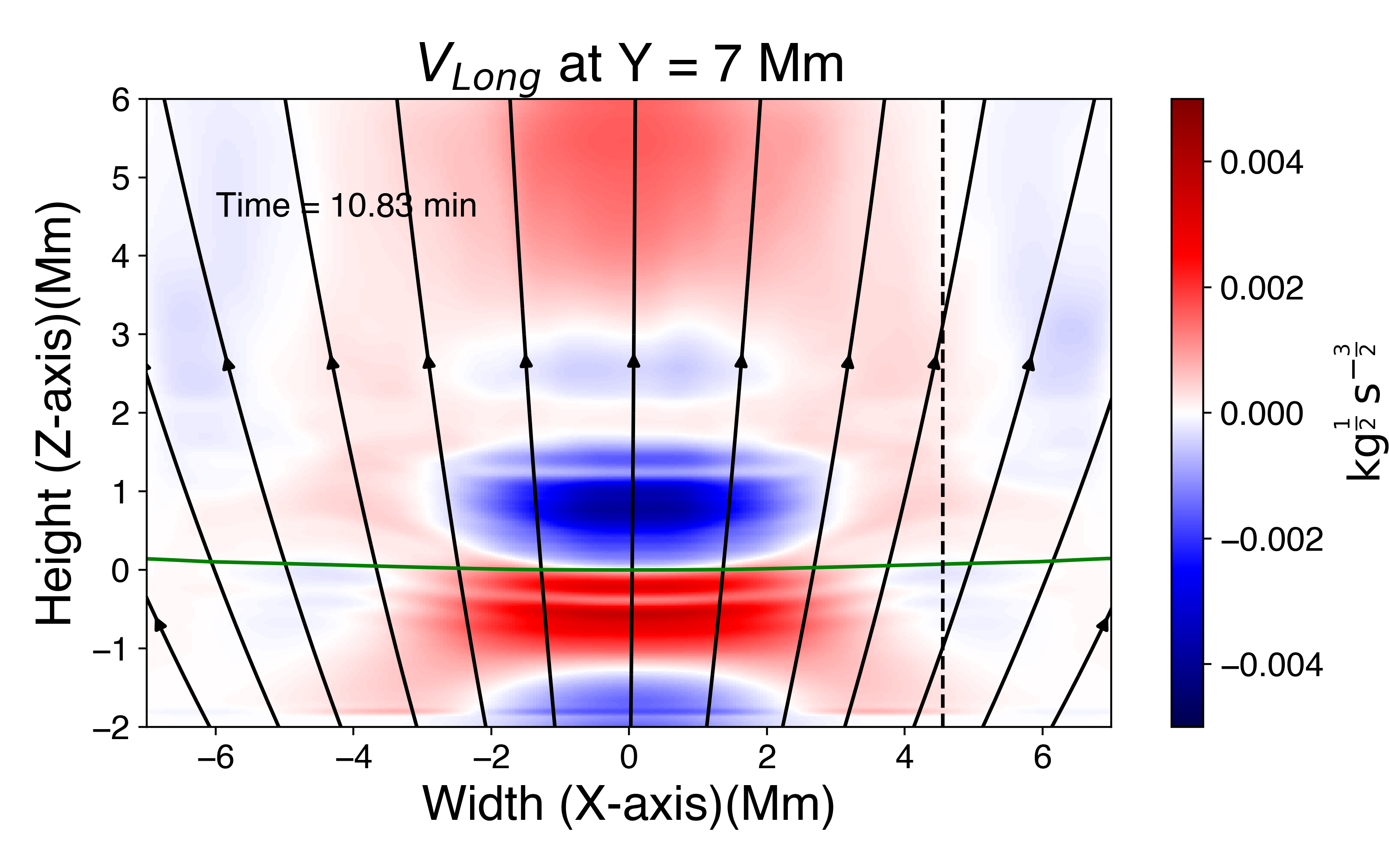}
    \includegraphics[width=0.5\textwidth]{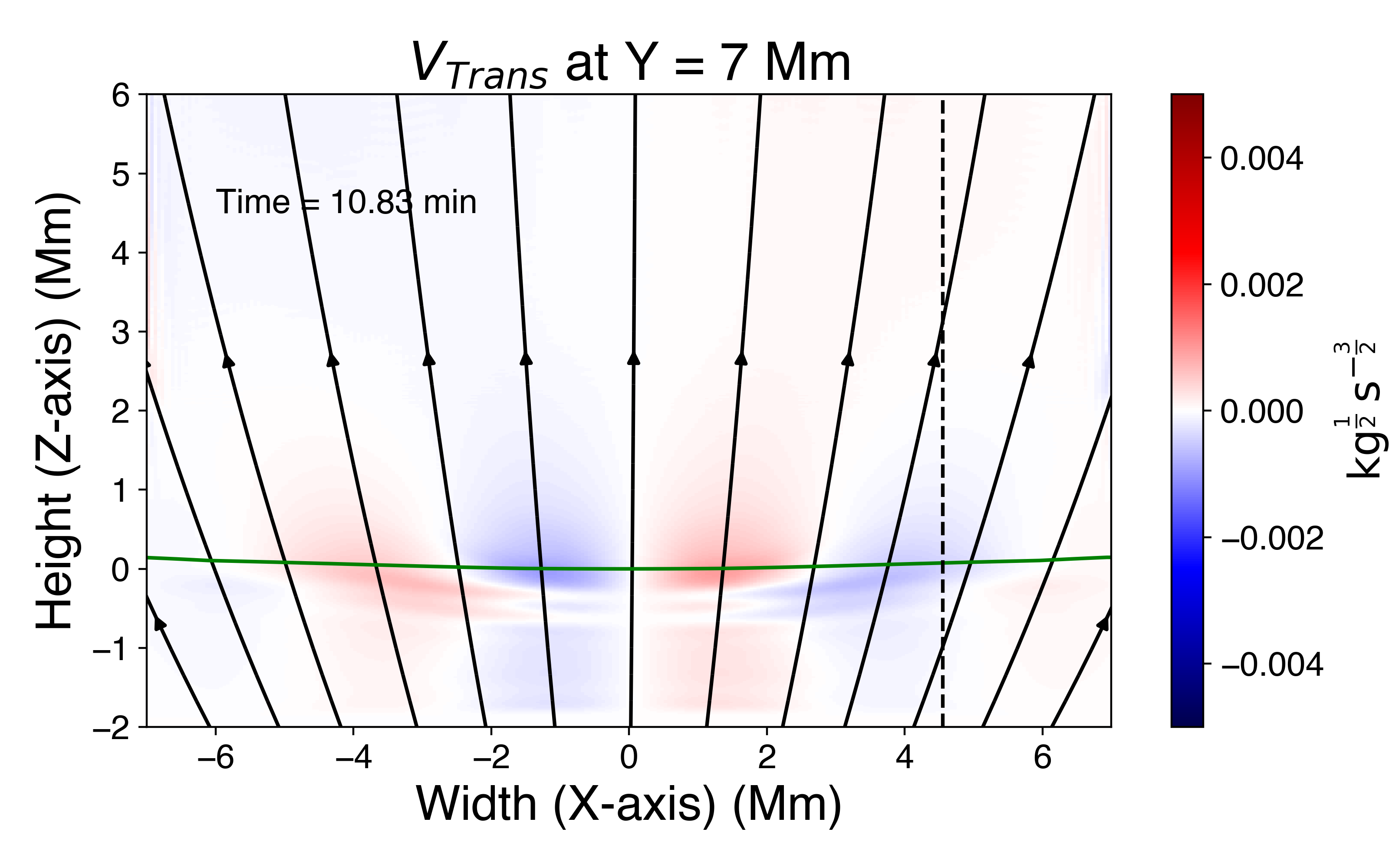}
    \includegraphics[width=0.5\textwidth]{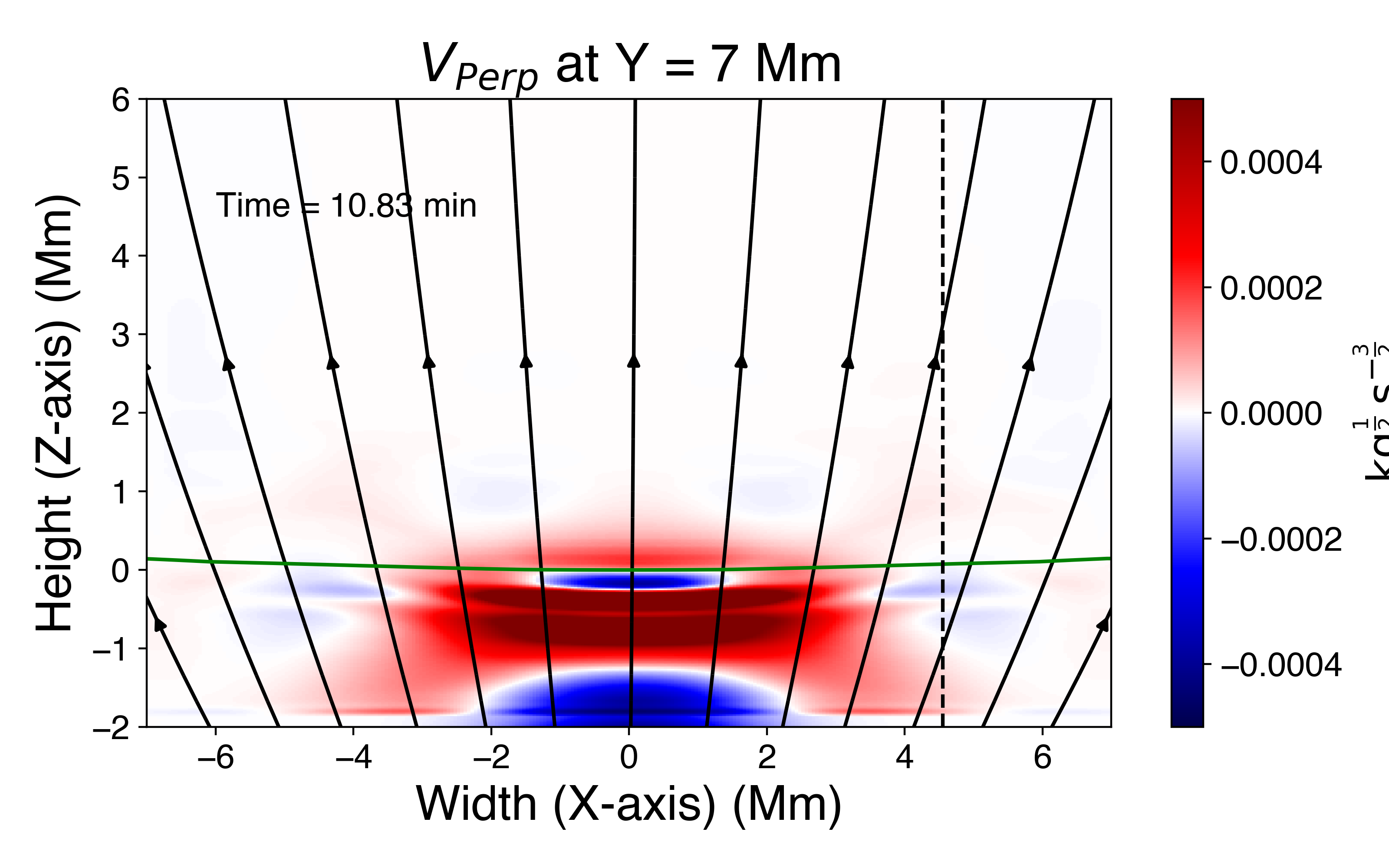}
    }       
    \caption{Left: Velocity projections of slow (left), fast (middle) and Alfv\'{e}n (right) waves at snapshot 1163 seconds. The green contour is the equipartition layer. The vertical dashed line is where the $\theta$ = $30^\circ$ and $\phi$ = $56^\circ$ at the equipartition layer. The velocities are scaled by a factor of $\sqrt{\rho_{0}c_{s}}$ on the left panel and $\sqrt{\rho_{0}v_{A}}$ for the middle and right panels.}
    \label{fig:modes_plane}
\end{figure*}

\begin{figure*}[t]
\centering
    \resizebox{\hsize}{!}{
    \includegraphics[width=0.5\textwidth]{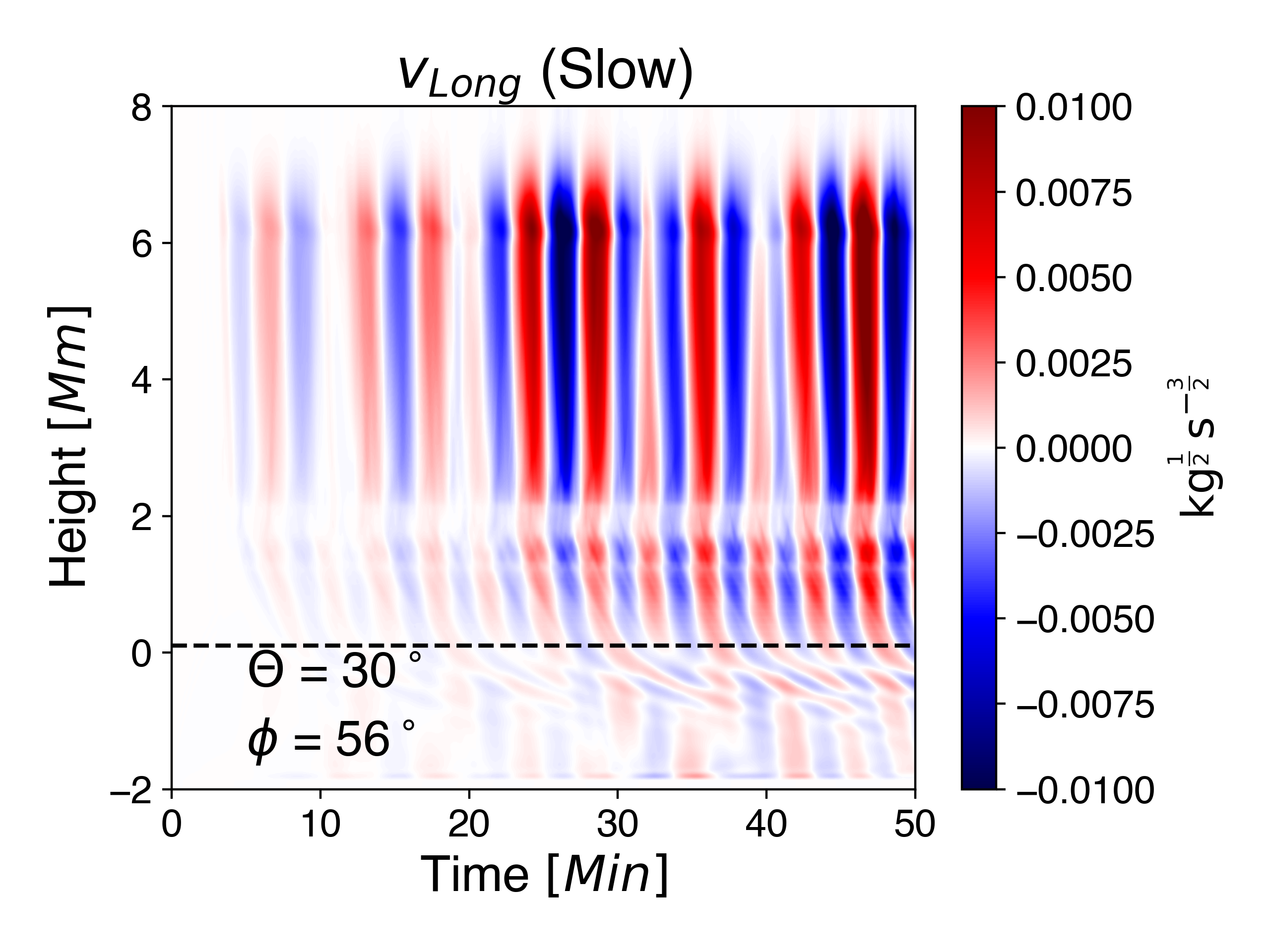}
    \includegraphics[width=0.5\textwidth]{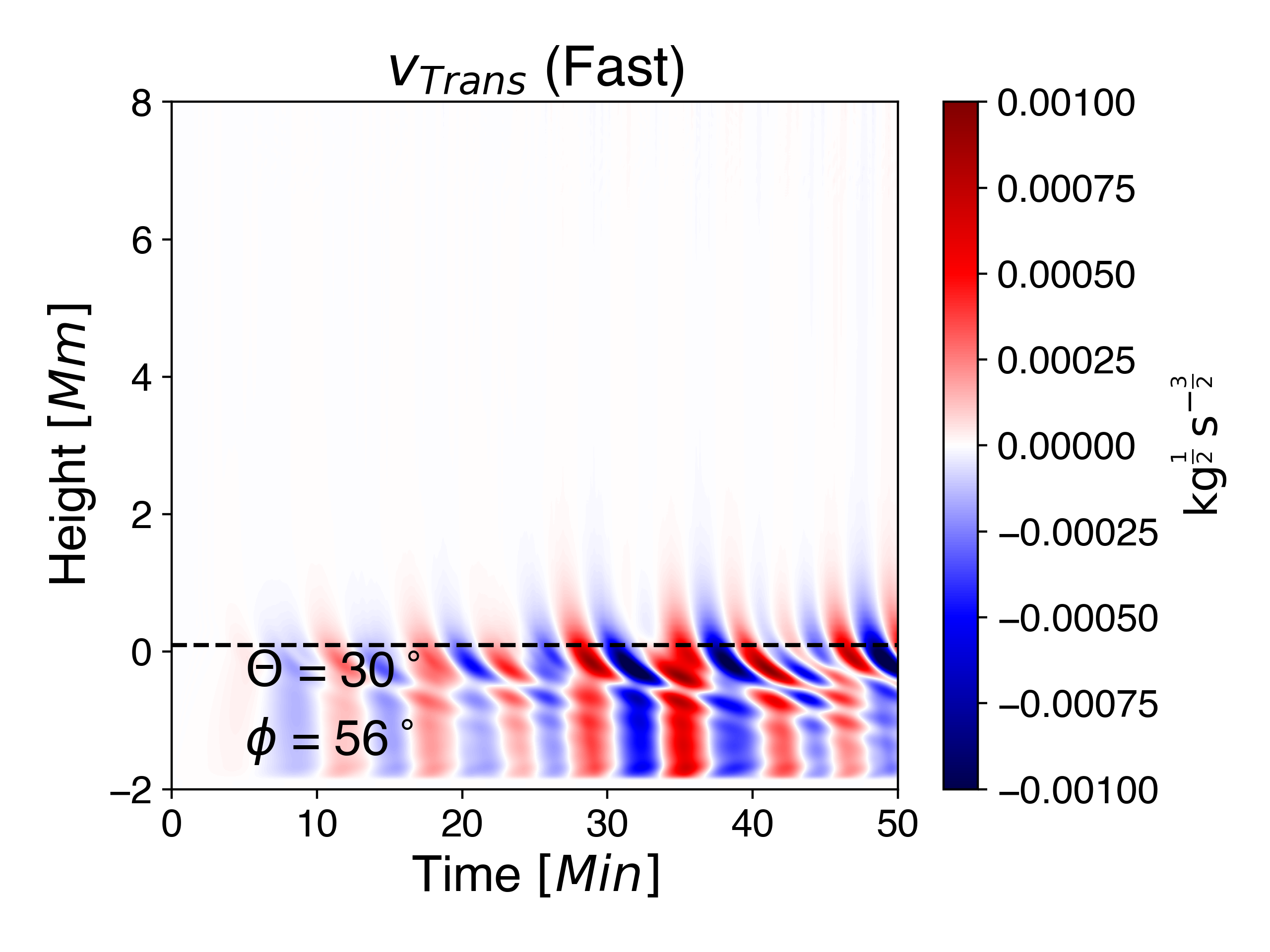}
    \includegraphics[width=0.5\textwidth]{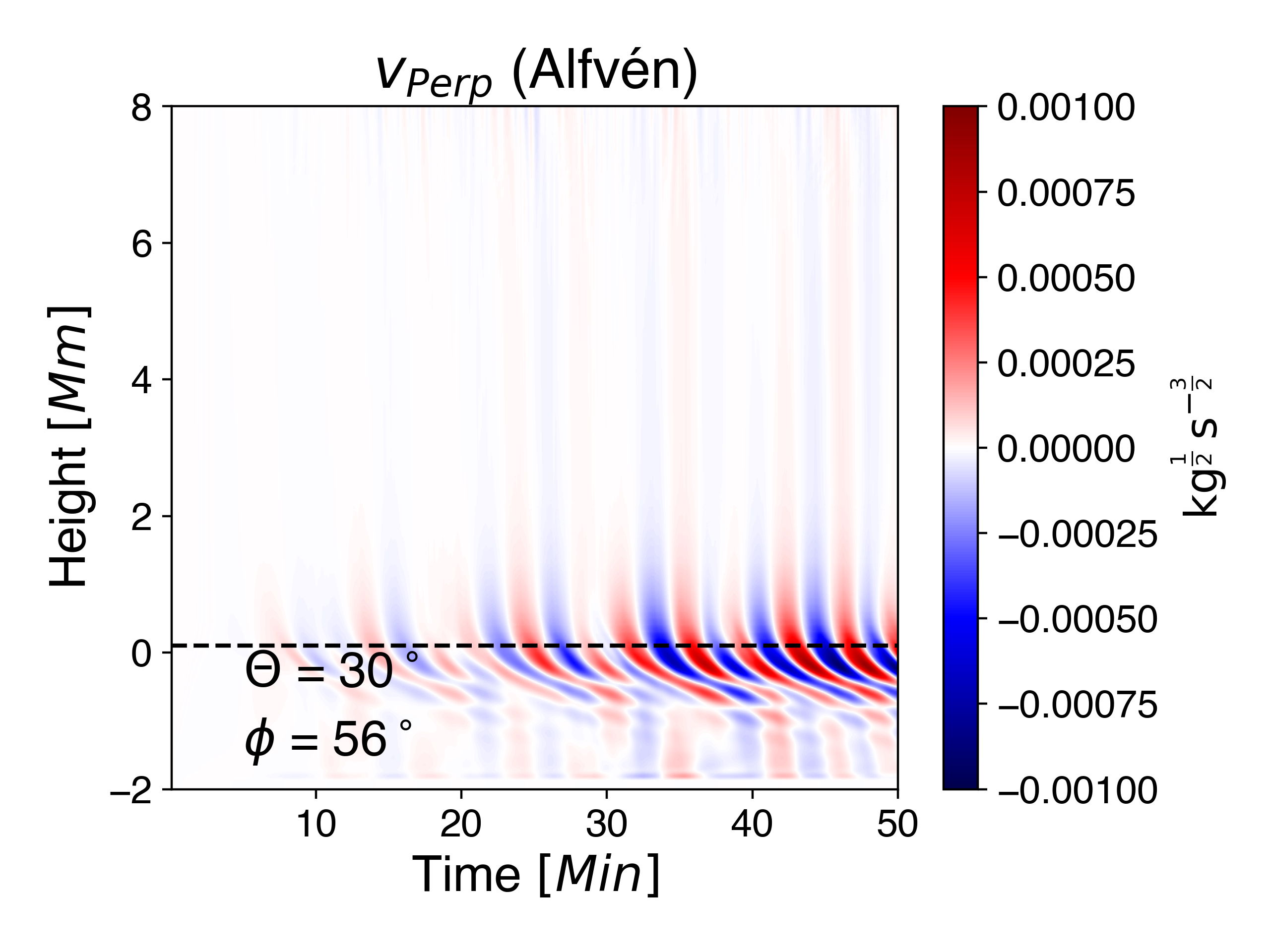}
    }  
    \resizebox{\hsize}{!}{
    \includegraphics[width=0.5\textwidth]{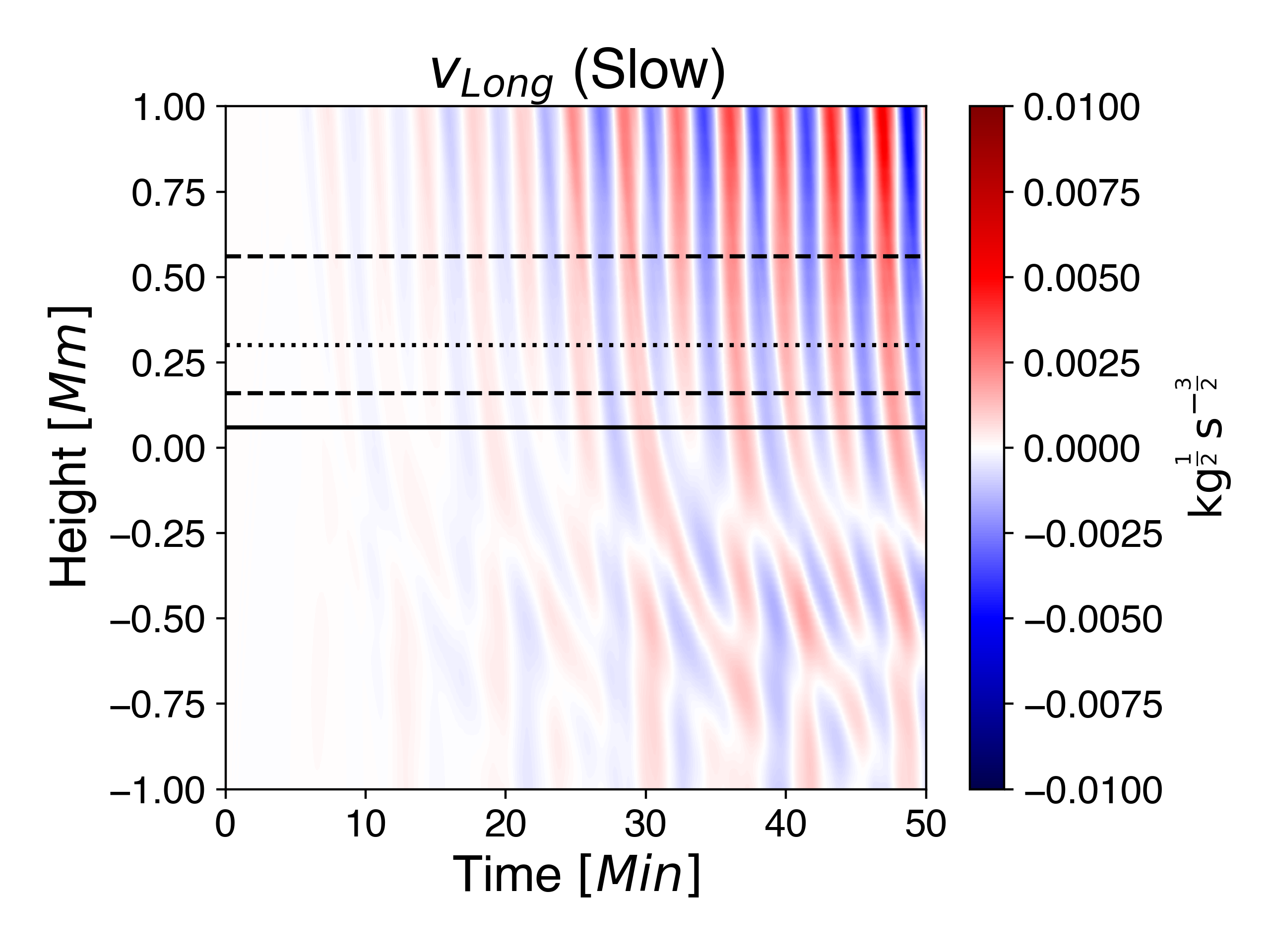}
    \includegraphics[width=0.5\textwidth]{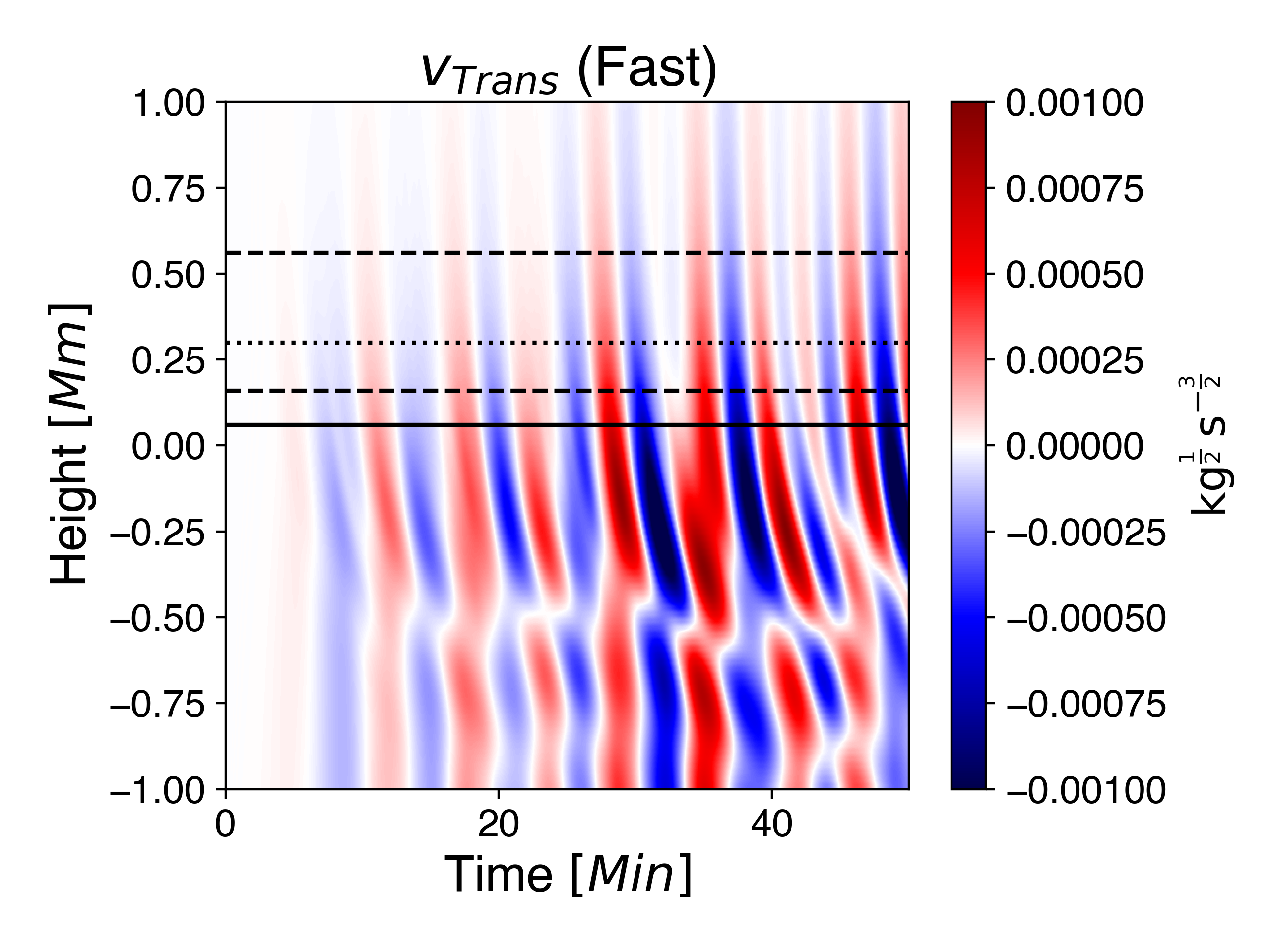}
    \includegraphics[width=0.5\textwidth]{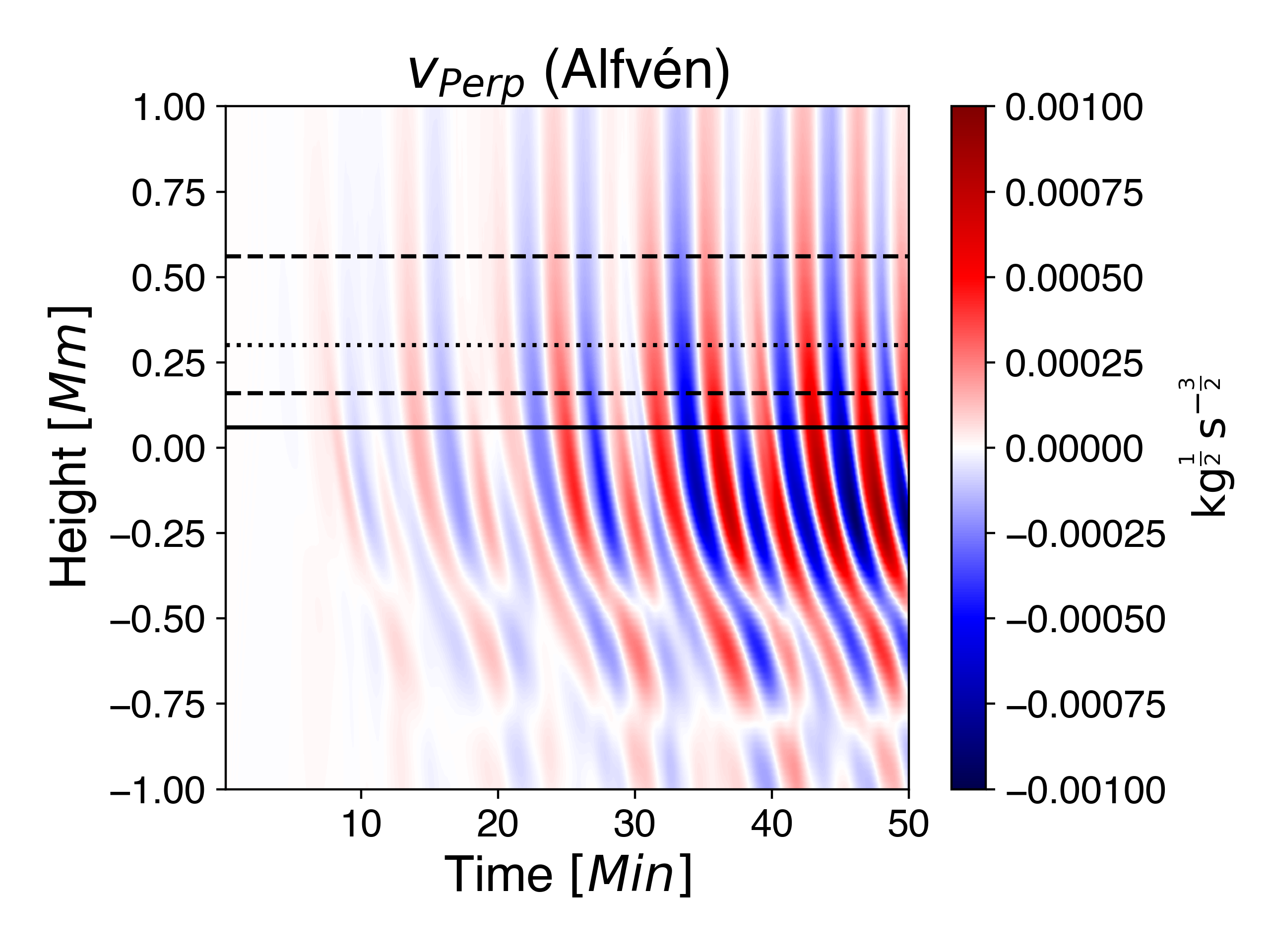}
    }
    \caption{Time evolution of velocity projections of slow (left), fast (middle) and Alfv\'{e}n (right) waves for a column of the simulation. The location in the domain is shown by the vertical line in Figure~\ref{fig:modes_plane}. The inclination angle and the azimuthal angles are $(\theta,\phi) = (30^\circ,56^\circ)$ at the equipartition layer. The plots in the second row are similar but zoomed to a vertical range of -1~Mm to 1~Mm for better clarity. The solid line is the equipartition layer and the region between the two dashed lines corresponds to the the fast wave reflection region. The dotted line is the the fast wave reflection region which is estimated by taking period $P_n=P_c$. All the velocity projections are scaled by a factor of $\sqrt{\rho_{0}c_{s}}$ on the left panel and $\sqrt{\rho_{0}v_{A}}$ for the middle and right panels.}
    \label{fig:modes_xt}
\end{figure*}

\section{Results}
Upon providing the background atmosphere and the driver, MANCHA is then used to solve the perturbations of density, pressure, magnetic field, and velocities of the system of MHD equations (Eqs.~\ref{eq:mass_conv}-\ref{eq:E}). Given that the mode conversions of interest are linear in nature, we keep the initial perturbations of the system small (refer Eqn.~\ref{eq:v_amp_driver}) to ensure the waves remain in the linear regime. We now discuss the results of the simulation in detail.
\pagebreak
\subsection{MHD Wave projections}
To differentiate the Alfv\'{e}n modes from the fast and slow magnetoacoustic modes in the magnetically dominated atmosphere ($v_{A}>c_s$), we use the following projections onto three characteristic directions:
\begin{eqnarray}\label{eq:proj}
    \hat{e}_{long} &=& [\cos\phi\sin\theta, \sin\phi\sin\theta, \cos\theta ]\\   
    \hat{e}_{perp} &=& [-\cos\phi\sin^{2}\theta\sin\phi,
    1 - \sin^{2}\theta \sin^{2}\phi, \nonumber \\
     &&-\cos\theta\sin\theta\sin\phi ] \label{eq:perp}\\ 
    \hat{e}_{trans} &=& [-\cos\theta,0, \cos\phi\sin\theta]. 
\end{eqnarray}
Here $\theta$ is the inclination of the magnetic field with respect to the vertical and $\phi$ is the azimuthal angle.
The slow ($long$) projection is the radial unit vector along the magnetic field. The Alfv\'{e}n ($perp$) projection is the asymptotic polarization direction perpendicular to the magnetic field as suggested by \citet{Cally2008}. The fast ($trans$) component is the cross product of slow and Alfv\'{e}n projections. These projections have been demonstrated to be rather effective in separating the perturbations related to all three modes \citep{Felipe2010, Khomenko2011}. Note that below the equipartition layer ($c_s>v_A$), the longitudinal and transverse components will contain a mixture of fast and slow perturbations, as the fast mode propagates isotropically.
 
An example of applying the wave projections to the velocity perturbations is shown in Figure~\ref{fig:modes_plane}. For visualisation, the ${long}$ projection is scaled by $\sqrt{\rho_0 c_s}$ and the ${trans}$ and ${perp}$ projections are scaled by $\sqrt{\rho_0 v_A}$. The scalings correspond to the respective wave energy fluxes,
    \begin{equation}
        F_w = \rho \langle v^2 \rangle v_{\rm ph}.
    \end{equation}
Here $\langle v^2 \rangle $ is the root mean square of velocity amplitudes and $v_{ph}$ is the phase speed. The magnitudes of the quantities in Figures \ref{fig:modes_plane}, \ref{fig:modes_xt} are then the square root of the kinetic energy flux. As the amplitude increases with height and density drops accordingly, the scaling factor $\sqrt{\rho_0v_{\rm ph}}$ allows for efficient visualisation of the respective velocity projections along different layers, reducing the effects of wave amplification with height due to the variation in plasma parameters. 

\subsection{Wave behaviour}

Figures~\ref{fig:modes_plane} and \ref{fig:modes_xt} demonstrate that the wave behaviour in the simulation is complex. However, the results for the lower solar atmosphere are comparable to the numerical results from the 3D simulations of \cite{Felipe2010,Khomenko2012}. 
From Figure~\ref{fig:modes_xt}, it is clear that both slow magnetoacoustic and Alfv\'en modes are able to propagate into the corona. The upward propagation of MHD waves from the lower solar atmosphere into the corona has been demonstrated in a number of previous numerical simulations, e.g., slow modes \citep{Hansten_2002, Botha_2011} and Alfv\'en modes \citep{Khomenko2019}. In contrast, there is little signature of the fast mode energy present. This is because the fast mode suffers significant reflection in a region above the equipartition layer, which is clearly observed in the middle panel of Figure~\ref{fig:modes_plane}. Above this region, the modes are evanescent in the corona \citep[e.g.,][]{Hollweg_1978b, Leroy_1982, Schwartz_1982}.

\subsubsection{Acoustic modes}
\begin{figure*}[t]
    \centering
    \resizebox{\hsize}{!}{
    \includegraphics[scale=1.5]{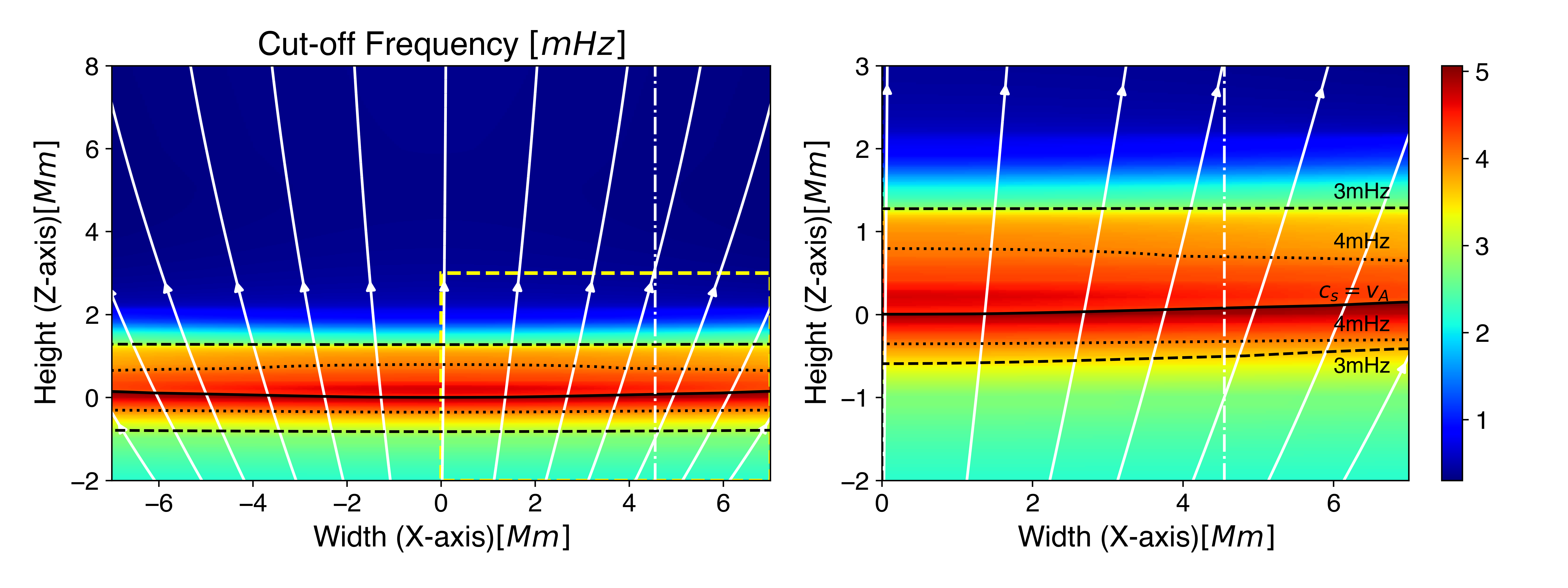}
    }
    \large
    \caption{Cutoff frequency across the background condition. White contours refer to the magnetic field lines. Vertical white dashdot line refers to the slit we considered to achieve the results in Figure \ref{fig:modes_xt}. The solid black line across the domain is the equipartition layer ($c_s = v_A$). The dashed and dotted lines refer to the cutoff layers where $v_{ac}$ = 4~mHz and 3~mHz respectively. We clipped the regions within the yellow dashed rectangular box to achieve the plot on right for better visualization purposes.}
    \label{fig:cutoff_backround}
\end{figure*}

Throughout the atmosphere, the acoustic modes are subject to reflections. For slow magnetoacoustic waves in a strongly magnetised environment ($c_s<v_A$), 
there are then two sources of wave reflection present, the gravitational stratification and also from regions with significant gradients in the pressure scale height \citep[ e.g.,][]{Roberts_2006,Botha_2011}. Figure~\ref{fig:cutoff_backround} displays the cut-off frequency arising from gravitational stratification (given by Eq.~\ref{eq:cutoff}) and it varies across the domain, peaking at $\sim4.5$~mHz. In the current model, slow waves with frequencies less than $\sim4$~mHz should be reflected before they reach the equipartition layer. 

Figure~\ref{fig:modes_plane} shows a number of locations where there is substantial wave reflection due to strong gradients in the pressure scale height ($H_p$), see Figures~\ref{fig:2Dcylinder} and \ref{fig:background_parameters} for sound speed profiles. These are visible as the horizontal stripes across the domain, notably at a height of $\sim2$~Mm in the slow projection. We expect this boundary to form a resonance cavity in the lower atmosphere, with the potential for standing modes to exist \citep[e.g.,][]{Zhugzhda, Zhugzhda_2008, Botha_2011, felipe_2020}.  Another horizontal stripe at 6~Mm marks where the PML boundary conditions start and the diffusion profile reaches a value half its maximum (see Figure \ref{diff}). This combination of factors leads to an artificial reflection point. Above 6~Mm the waves are damped rapidly showing the effectiveness of the combined PML and artificial diffusion.

\medskip 

It is insightful to examine how the waves evolve with time in the system. Figure~\ref{fig:modes_xt} shows the velocity projections for a single column of the simulation (its location in the domain at X= 4.55~Mm, marked by the vertical dashed lines in Figure~\ref{fig:modes_plane}). 
For this column, the magnetic field inclination has values ($\theta, \phi$)=($30^\circ$, $56^\circ$) at the equipartition layer. It can be seen that the excited fast acoustic waves propagate upwards and are  either split into slow magnetoacoustic waves or mode converted to fast magnetoacoustic waves at the equipartition layer (indicated by the solid line running across all six panels). In the left panels of Figure~\ref{fig:modes_xt}, the propagating slow waves above the equipartition layer experience strong reflection from the locations with steep temperature gradients of the transition region. This feature was not seen in the simulations of \cite{Khomenko2012}, but is similar to 2D simulations of \citet{santamaria2015magnetohydrodynamic}. The returning slow magneto acoustic mode should then also be able to mode convert or be transmitted as they pass back through the equipartition layer. 

The upwardly travelling slow waves then propagate quickly through the coronal part of the simulation due to the increased sound speed. The slow waves suffer reflection around 6~Mm due the onset of the diffusion profile and PML layers. This leads to a variable flux of downward propagating slow modes, indicated by the varying slopes in the time-distance diagram of the longitudinal velocity component.

\subsubsection{Transverse modes}
In the middle panels of Figures~\ref{fig:modes_plane} and \ref{fig:modes_xt}, the fast magnetoacoustic wave is chosen by the orthogonal projection $\hat{e}_{trans}$ above the equiparition layer, but it is a mixture of wave modes below the equipartition layer.
The enhancement of the transverse component about the equipartition is dominated by downward propagating, reflected modes (Figure~\ref{fig:modes_xt} middle panels).  The reflection of fast magnetoacoustic waves back into the lower atmosphere occurs up to around 2~Mm above the equipartition layer. The region of wave reflection will differ for modes with different $\omega$ and $k_h$. In Figure~\ref{fig:modes_xt} we indicate the reflection region for a wave with $k_h=1.07$~Mm$^{-1}$ (equivalent to full-width half-maximum of driver pulse). The lower dashed line refers to the height where $\omega = 2\pi/P_n$, for $P_n = 100$ seconds and the upper dashed line refers to the region where $P_n = 600$ seconds. The dotted line refers to the region where $P_n = P_c$, i.e. 320~s where the driver is centred. As the fast to fast conversion is linear, the fast magnetoacoustic modes generated in this simulation should be reflected strongly below 1~Mm. This is clearly seen in the time-distance plots. The observed pattern of reflection is comparable to that observed in previous simulations \citep[e.g.,][]{Felipe2010, Khomenko2012}. As mentioned, very little fast magnetoacoustic (or transverse) energy is able to reach the corona. 

Due to the presence of multiple wave frequencies and reflection of slow and Alfv\'en waves from the transition region, the pattern below $\sim-1$~Mm is more complicated than previous simulations. This is because the mode conversion is possible between all modes in this region \citep{Cally_2021}. Although interesting, we do not attempt to disentangle the relationships between the reflected waves.

\subsubsection{Alfv\'en modes}
The Alfv\'{e}n waves are separated by the projection $\hat{e}_{perp}$ from Eq.~(\ref{eq:perp}). 
Given the strong reflection of the transverse wave modes observed within Figure~\ref{fig:modes_xt}, it can be expected that some of the fast wave energy is converted to upwardly propagating Alfv\'en waves, hence the occurrence of the perpendicular component in the corona \citep[e.g., also found in the simulations of][]{Felipe2010,Khomenko2012}. The Alfv\'en waves are reflected throughout the simulation due to gradients in the Alfv\'en speed \citep[e.g.,][]{Hollweg1978,Schwartz_1984}, and some of the wave energy is reflected back towards the photosphere. The steepest gradients in Alfv\'en speed occur at the transition region (see, Figures~\ref{fig:2Dcylinder} and \ref{fig:background_parameters}). The reflected Alfv\'{e}n waves leave a significant signature of downward propagation in the right panels of Figure~\ref{fig:modes_xt}.  This reflection reduces the amount of Alfv\'en wave energy able to reach the upper part of simulation \citep[compared to the results of][]{Khomenko2012}.

Hence, irrespective of the wave types, the transition region acts as a partial barrier to all the upward propagating waves within this simulation. It has been shown that transverse structuring can aid the transmission of waves to the corona \citep{Khomenko2019,skirvin}, although there is a suggestion that the rate of expansion of the magnetic field in the lower atmosphere is the dominant influence on wave energy flux through the transition region \citep[at least for Alfv\'en waves;][]{Taroyan_2024}. Here, the sunspot is best described as a thick flux tube model that rapidly expands in the lower atmosphere and this leads to strong reflection.

\begin{figure*}[t] 
    \centering
    \resizebox{\hsize}{!}{
    \includegraphics[scale=0.1]{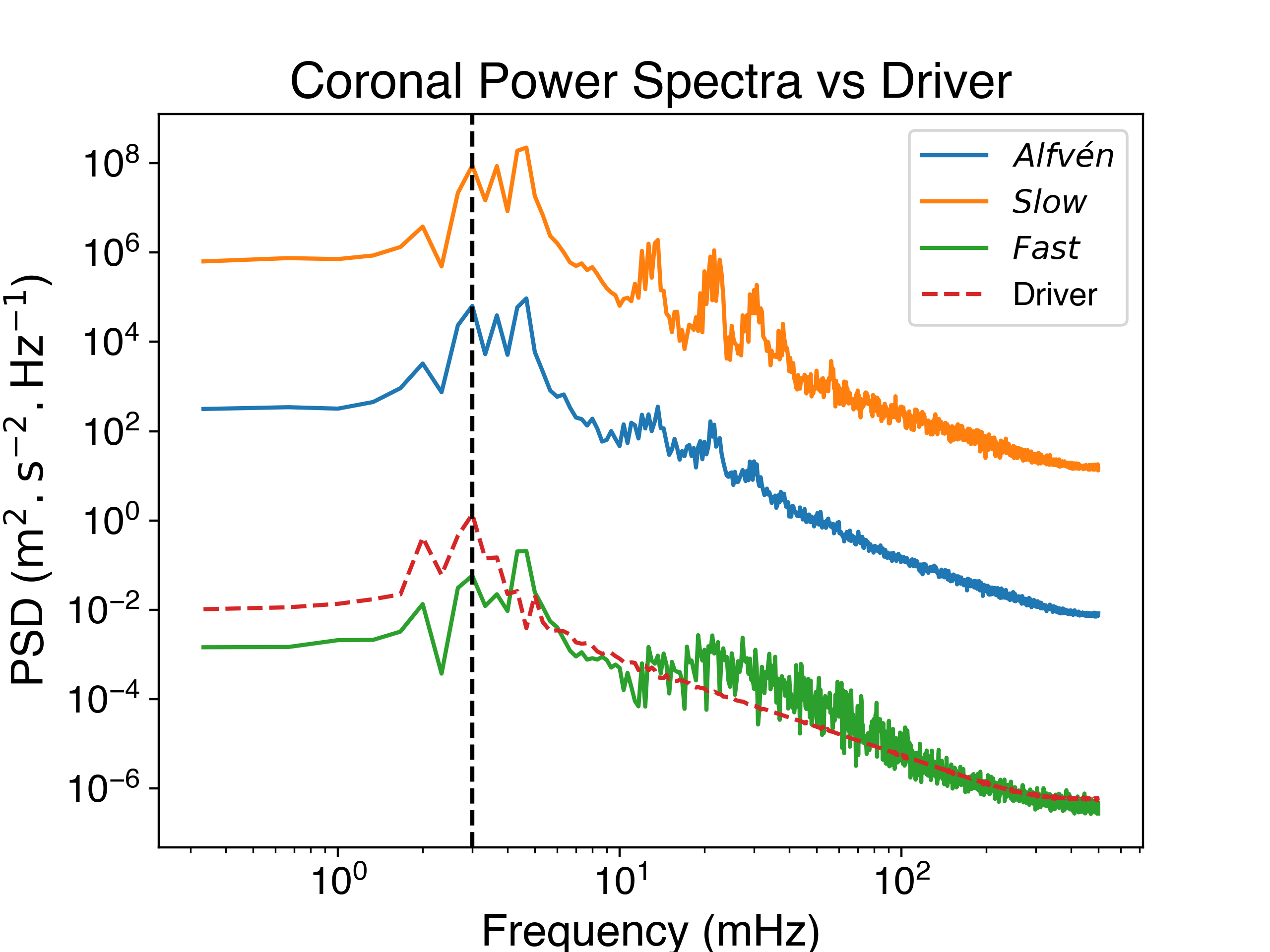} 
    \includegraphics[scale=0.1]{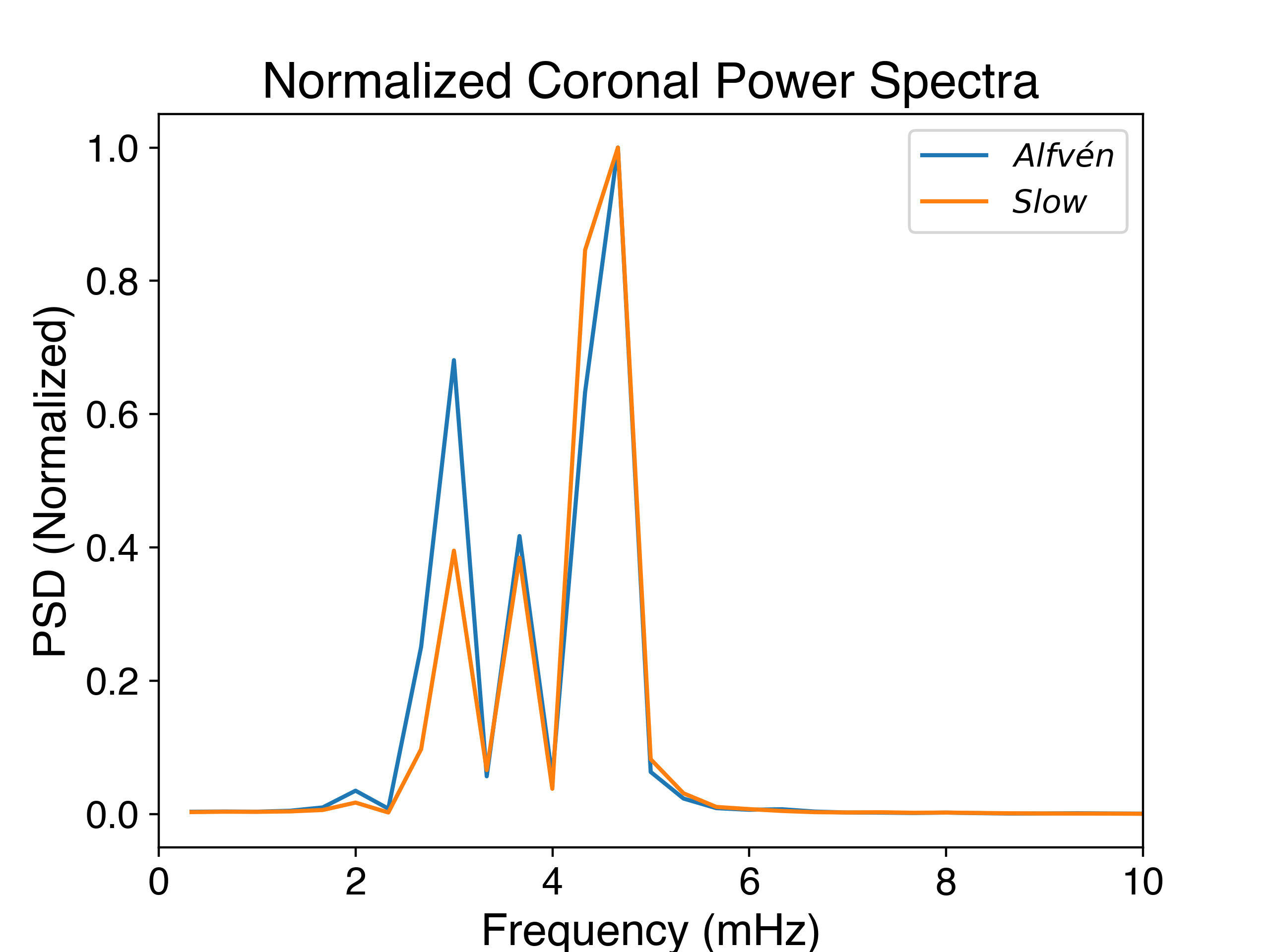}
    }
    \large
    \caption{ Coronal power spectrum for the MHD wave modes. The figure on the left shows average power spectra for  slow, Alfv\'{e}n and fast velocity projections, indicated by the the orange, blue and green curves, respectively. For all modes, the coronal Alfv\'{e}n power spectrum peaks around 4.5~mHz, with the frequency indicated by the vertical black dotted line. The red dashed curve is the averaged power spectrum of velocity projection $v_z$ at the location of the driver. The peak frequency of the driver is shown by the vertical black dashed line. The figure on the right is the  normalised average power spectra for slow, Alfv\'{e}n projections.}
    \label{fig:power}
\end{figure*}

\subsection{Coronal Power Spectra}
To determine the coronal power spectra of different waves modes, we take the Fourier transform  
of the individual time series at each grid point for each velocity projection. The time-series used are taken across the sunspot domain at the height of $Z=4$~Mm. The squared absolute value of the Fourier coefficients is taken and averaged across all time-series to obtain the average power spectra across the sunspot. The averaging across the sunspot will somewhat mimic line-of-sight integration through the corona at the limb, incorporating the wave behaviour across magnetic fields with various inclination angles. 

Figure \ref{fig:power} shows the coronal power spectra of fast, slow and Alfv\'{e}n waves. In addition to these curves, we also show the power spectrum of velocity projection $v_z$, averaged across the width of the driving pulse ($\sigma_x$) at the height of $Z=-1.65$~Mm. It can be seen that the coronal power spectra for all the wave modes have an enhanced power which is clearly located at different frequency from that of the driver (with dominant driving frequency at $\approx3$~mHz). Fitting a Gaussian function to the power enhancement for the Alfv\'en waves reveals the peak occurs at a frequency of $\sim4$~mHz. The frequency is comparable to that estimated from the coronal Doppler velocity fluctuations associated with Alfv\'enic waves \citep{morton2019}. We note that the high-frequency peaks in the coronal power spectra between 10-50~mHz are caused by the spurious excitation of waves due to the reflection from the periodic boundaries.

As discussed in the introduction, this phenomenon arises due to frequency filtering effects present for magneto-acoustic wave propagation. The upward propagating acoustic modes are subject to frequency dependent reflection below the equipartition layer, leading to a filtering. The characteristics of the acoustic wave power spectra are then passed on to the other wave modes during the linear mode conversion processes. This is indicated by the fact that all the coronal power spectra display an enhancement of power in the same frequency range (and have similarly shaped power spectra).  

For the current simulation, the coronal slow modes have substantially greater power than the coronal Alfv\'en modes. Although, as discussed, there is a nonphysical reflection of the coronal slow waves due to the numerical implementation which means the magnitude of the power difference between slow and Alfv\'en modes is likely inflated. The transverse waves have a factor of $\sim10^5$ less power than the Alfv\'en waves, which is expected due to the near total reflection of fast modes.

\section{Conclusion and discussion}
It is well established that the coronal Alfv\'{e}nic power spectrum has an enhancement around 4~mHz \citep{Tomczyk2007, morton2019}, and the underlying cause behind the enhanced power at this frequency range has been the subject of debate. Previous work has suggested that \textit{p}-modes (which have a peak power of around 3~mHz) could be responsible, mode converting first to fast magnetic waves then to Alfv\'en waves, \cite[e.g.,][]{Cally2011, Hansen_2012, Khomenko2012, Cally2016}. Here, we demonstrate by direct numerical simulation that the enhancement of the coronal Alfv\'{e}nic power spectrum can be directly connected to the internal acoustic oscillations.

As discussed in the introduction, the temperature structure of the lower solar atmosphere is a natural filter for the upward propagating acoustic waves, truncating the power spectra for low frequencies ($\nu< \nu_{ac}$). The subsequent mode conversions, i.e., fast to fast, and fast to Alfv\'en, are linear so the filtered power spectrum of the \textit{p}-modes is imparted on the resultant transformed wave modes. This is clearly seen in the simulations when measuring the coronal power spectrum for all wave modes. The peak of the coronal power spectrum will depend upon the location of equipartition layer with respect to the temperature minimum (where the highest value frequency cutoff occurs). In the current model, the equipartition layer is close to, but below, the temperature minimum, hence the upwardly propagating waves are subject to nearly the maximal filtering. Should the equiparition layer be lower, then the filtering of the \textit{p}-mode spectrum would be less and the Alfv\'enic power spectrum would likely peak at a lower frequency.

The slow magnetoacoustic modes continue to feel the effect of cutoff above the equipartition layer, and should be more strongly filtered. This is seen when comparing the normalised coronal power spectrum for the slow and Alfv\'en waves (Figure~\ref{fig:power} right panel), where the slow waves have less power at lower frequencies.
The magnetic field inclination ($\theta$) will also play a role in shaping the coronal wave power spectrum as it further modifies the cutoff frequency. This will essentially only effect the slow modes though.

\medskip 

We note that in the current simulation, there is a limited range of magnetic field inclination angles (30$^\circ$ $<\theta<$ 40$^\circ$). The range of angles is particularly suited to the fast-to-Alfv\'en conversion. Further simulations are required that incorporate a broader selection of magnetic field inclinations (both $\theta$ and $\phi$) to see whether the peak from the average coronal power spectra still occurs at 4~mHz. We speculate that this will be the case as the coronal Alfv\'en spectrum is shaped by the filtering of the fast acoustic modes below the equipartition layer, which is largely independent of inclination.

The results from the sunspot should also be representative of wave dynamics in network fields in the quiet Sun. The foreseeable difference is the relative heights of the equipartition layer and the temperature minimum. In network elements the equipartition layer is likely to occur above the temperature minimum \citep[e.g., see atmospheric structure of network element in][]{Khomenko2008b}, although this will depend upon the magnetic field strength. This might introduce additional filtering of the fast acoustic modes before they are converted to fast magnetoacoustic modes. However, the minimum plasma temperature in network elements is likely greater than that in the sunspots (potentially be up to 1000~K more based on 1D semi-empirical models of sunspots and network elements). This means the frequency filtering could be less severe and may not extend to frequencies of 5~mHz. Further simulations would be required to confirm this.

It is worth highlighting that we derive an averaged coronal Alfvén power spectrum at 4~Mm above the photosphere, far below the heights at which CoMP has previously made measurements. We believe that the spectrum would remain largely unchanged as the waves propagate higher in the corona. These Alfv\'en waves are subject to amplification with height due to the drop in density with height and observations suggest they remain linear. Furthermore, observations indicate there is weak Alfv\'enic wave damping in the quiet Sun \citep{tiwari_2021} and coronal holes \citep{morton_2015} from frequency-dependent mechanisms such as resonant absorption and phase mixing. Hence, we expect the overall shape of the power spectrum to remain largely unchanged.

\begin{acknowledgements}
H. M. thanks Northumbria University for financial support. R. J. M. would like to thank the UKRI for financial support via a UKRI Future Leader Fellowship
(RiPSAW MR/T019891/1). E.K. is grateful for the support by the Spanish Ministry of Science and Innovation through the grant PID2021-127487NB-I00. The authors would like to thank K. Karampelas, T. Duckenfield and R. Sharma for reading drafts and valuable discussions. 
\end{acknowledgements}

\bibliography{bibliography1}{}
\bibliographystyle{aasjournal}
\end{document}